\documentclass[twocolumn,tighten,trackchanges]{aastex62}
\pdfoutput=1
\usepackage{amsmath,amstext,amssymb}
\usepackage[T1]{fontenc} 
\usepackage{apjfonts} 
\usepackage[figure,figure*]{hypcap}
\usepackage{comment}
\usepackage{subfigure}
\usepackage{mwe}
\usepackage{xcolor}
\usepackage[ampersand]{easylist}
\usepackage{multirow}
\usepackage{booktabs}
\usepackage{wrapfig}
\usepackage[document]{ragged2e}
\usepackage{epstopdf}
\usepackage{pgffor}
\usepackage{hyperref}
\usepackage{gensymb}
\usepackage[switch, modulo]{lineno} 
\usepackage{enumerate} 
\usepackage[ampersand]{easylist}
\usepackage[colorinlistoftodos]{todonotes}


 
\shorttitle{A NEW CLASS OF CHANGING-LOOK LINERS}
\shortauthors{Frederick et al.}

\newcommand{\myemail}{sfrederick@astro.umd.edu}
\newcommand{\tyrion}{ZTF18aajupnt/AT2018dyk}
\newcommand{\bronn}{ZTF18aahiqfi}
\newcommand{\varys}{ZTF18aaidlyq}
\newcommand{\noname}{ZTF18aasuray}
\newcommand{\podrick}{ZTF18aasszwr}
\newcommand{\jorah}{ZTF18aaabltn}
\newcommand{\bco}{iPTF16bco}
\newcommand{\clagn}{changing-look AGN}
\newcommand{\num}{six}

\emergencystretch=\maxdimen
\accepted{August 8, 2019}

\begin{document}
\justify
\title{A NEW CLASS OF CHANGING-LOOK LINERS}

\correspondingauthor{Sara Frederick}
\email{\myemail}

\author[0000-0001-9676-730X]{Sara Frederick}
\affiliation{Department of Astronomy, University of Maryland, College Park, MD 20742, USA}

\author[0000-0003-3703-5154]{Suvi Gezari}
\affiliation{Department of Astronomy, University of Maryland, College Park, MD  20742, USA}
\affiliation{Joint Space-Science Institute, University of Maryland, College Park, MD 20742, USA}

\author[0000-0002-3168-0139]{Matthew J. Graham}
\affiliation{Division of Physics, Mathematics, and Astronomy, California Institute of Technology, Pasadena, CA 91125, USA}

\author[0000-0003-1673-970X]{S. Bradley Cenko}
\affiliation{Joint Space-Science Institute, University of Maryland, College Park, MD 20742, USA}
\affiliation{Astrophysics Science Division, NASA Goddard Space Flight Center, MC 661, Greenbelt, MD 20771, USA}

\author[0000-0002-3859-8074]{Sjoert van Velzen}
\affiliation{Department of Astronomy, University of Maryland, College Park, MD 20742}
\affiliation{Center for Cosmology and Particle Physics, New York University, NY 10003}

\author[0000-0003-2686-9241]{Daniel Stern}
\affiliation{Jet Propulsion Laboratory, California Institute of Technology, 4800 Oak Grove Drive, Mail Stop 169-221, Pasadena, CA 91109, USA}

\author[0000-0003-0901-1606]{Nadejda Blagorodnova}
\affiliation{Department of Astrophysics/IMAPP, Radboud University, Nijmegen, The Netherlands}

\author[0000-0001-5390-8563]{Shrinivas R. Kulkarni}
\affiliation{Division of Physics, Mathematics, and Astronomy, California Institute of Technology, Pasadena, CA 91125, USA}

\author[0000-0003-1710-9339]{Lin Yan}
\affiliation{Caltech Optical Observatories, California Institute of Technology, Pasadena, CA 91125, USA}


\author[0000-0002-8989-0542]{Kishalay De}
\affiliation{Division of Physics, Mathematics, and Astronomy, California Institute of Technology, Pasadena, CA 91125, USA}

\author{U. Christoffer Fremling}
\affiliation{Division of Physics, Mathematics, and Astronomy, California Institute of Technology, Pasadena, CA 91125, USA}

\author[0000-0002-9878-7889]{Tiara Hung}
\affiliation{Department of Astronomy and Astrophysics, University of
California, Santa Cruz, CA 95064, USA}

\author[0000-0003-0172-0854]{Erin Kara}
\affiliation{Department of Astronomy, University of Maryland, College Park, MD 20742, USA}
\affiliation{Joint Space-Science Institute, University of Maryland, College Park, MD 20742, USA}
\affiliation{X-ray Astrophysics Laboratory, NASA/Goddard Space Flight Center, Greenbelt, Maryland 20771, USA} 

\author{David L. Shupe}
\affiliation{IPAC, California Institute of Technology, 1200 E. California Blvd, Pasadena, CA 91125, USA}

\author{Charlotte Ward}
\affiliation{Department of Astronomy, University of Maryland, College Park, MD 20742, USA}


\author[0000-0001-8018-5348]{Eric C. Bellm}
\affiliation{DIRAC Institute, Department of Astronomy, University of Washington, 3910 15th Avenue NE, Seattle, WA 98195, USA} 

\author{Richard Dekany}
\affiliation{Caltech Optical Observatories, California Institute of Technology, Pasadena, CA 91125, USA}

\author{Dmitry A. Duev}
\affiliation{Division of Physics, Mathematics, and Astronomy, California Institute of Technology, Pasadena, CA 91125, USA}

\author{Ulrich Feindt}
\affiliation{The Oskar Klein Centre, Physics Department, Stockholm University, Albanova University Center, SE 106 91 Stockholm, Sweden}

\author{Matteo Giomi}
\affiliation{Institute of Physics, Humboldt Universit{\ae}t zu Berlin, Newtonstra{\ss}e 15, 12489 Berlin, Germany} 


\author{Thomas Kupfer}
\affiliation{Kavli Institute for Theoretical Physics, University of California, Santa Barbara, CA 93106, USA}

\author{Russ R. Laher}
\affiliation{IPAC, California Institute of Technology, 1200 E. California Blvd, Pasadena, CA 91125, USA}
             
\author{Frank J. Masci}
\affiliation{IPAC, California Institute of Technology, 1200 E. California Blvd, Pasadena, CA 91125, USA}

\author[0000-0001-9515-478X]{Adam A. Miller}
\affil{Center for Interdisciplinary Exploration and Research in Astrophysics (CIERA) and Department of Physics and Astronomy, Northwestern University, 2145 Sheridan Road, Evanston, IL 60208, USA}
\affil{The Adler Planetarium, Chicago, IL 60605, USA}

\author{{James D. Neill}}
\affiliation{Division of Physics, Mathematics, and Astronomy, California Institute of Technology, Pasadena, CA 91125, USA}
             
\author{Chow-Choong Ngeow}
\affiliation{Graduate Institution of Astronomy, National Central University,
Taoyuan 32001, Taiwan}

\author[0000-0002-4753-3387]{Maria T. Patterson}
\affiliation{DIRAC Institute, Department of Astronomy, University of Washington, 3910 15th Avenue NE, Seattle, WA 98195, USA}

\author{Michael Porter}
\affiliation{Caltech Optical Observatories, California Institute of Technology, Pasadena, CA 91125, USA}

\author{Ben Rusholme}
\affiliation{IPAC, California Institute of Technology, 1200 E. California Blvd, Pasadena, CA 91125, USA}

\author{Jesper Sollerman}
\affiliation{The Oskar Klein Centre \& Department of Astronomy, Stockholm University, AlbaNova, SE-106 91 Stockholm, Sweden}

\author{Richard Walters}
\affiliation{Caltech Optical Observatories, California Institute of Technology, Pasadena, CA 91125, USA}

\begin{abstract}
We report the discovery of six active galactic nuclei (AGN) caught ``turning on" during the first nine months of the Zwicky Transient Facility (ZTF) survey. 
The host galaxies were classified as LINERs by weak narrow forbidden line emission in their archival SDSS spectra, and detected by ZTF as nuclear transients. 
In five of the cases, we found via follow-up spectroscopy that they had transformed into broad-line AGN, reminiscent of the changing-look LINER \bco. 
 In one case, ZTF18aajupnt/AT2018dyk, follow-up {\it HST} UV and ground-based optical spectra revealed the transformation into a narrow-line Seyfert 1 (NLS1) with strong [Fe~VII, X, XIV] and He~II $\lambda 4686$ coronal lines. 
 {\it Swift} monitoring observations of this source reveal bright UV emission that tracks the optical flare, accompanied by a luminous soft X-ray flare that peaks $\sim$60 days later.  {\it Spitzer} follow-up observations also detect a luminous mid-infrared flare implying a large covering fraction of dust. 
 Archival light curves of the entire sample from CRTS, ATLAS, and ASAS-SN constrain the onset of the optical nuclear flaring from a prolonged quiescent state. 
Here we present the systematic selection and follow-up of this new class of changing-look LINERs, compare their properties to previously reported changing-look Seyfert galaxies, and conclude that they are a unique class of transients well-suited to test the uncertain physical processes associated with the LINER accretion state.
\end{abstract}

\section{Introduction}
\label{intro}
The observed diversity in the optical spectra of AGN, with well-defined systematic trends known as the eigenvector relations, are understood to be a function of both orientation as well as accretion rate (e.g. \citealt{Shen2014}).  ``Changing-look" active galactic nuclei (CLAGN) are a growing class of objects that are a challenge to the orientation-based unification picture, in that they demonstrate the appearance (or disappearance) of broad emission lines and a non-stellar continuum, changing their classification between type 1.8-2 (narrow-line) to type 1 (broad-line) AGN (or vice versa) on a timescale of years. The nature of this spectral transformation is most often attributed to changes in accretion rate \citep{Shappee2014,Runnoe2016,MacLeod2016,Oknyansky2016,Ruan2016, Sheng2017}, but the mechanism(s) driving these sudden changes is still not well understood (e.g. \citealt{Lawrence2018,Stern2018}). 

One of the known changing-look quasars (CLQs), \bco~ \citep{Gezari2017}, was caught ``turning-on" in the iPTF survey into a broad-line quasar from a low-ionization nuclear emission-line region galaxy (LINER).  LINERs are distinguished from Seyfert 2 (Sy~2) spectra via the relatively strong presence of low-ionization or neutral line emission from [O~I] $\lambda$6300, [O~II] $\lambda$3727, [N~II] $\lambda\lambda$6548, 6583, and [S~II] $\lambda\lambda$6717, 6731; a lower [O~III] $\lambda$5007/H$\beta$ flux ratio; and a lower nuclear luminosity. However, the status of LINERs as low-luminosity AGN remnants is a topic of debate, as weak emission in some LINERs could also be powered by shocks, winds, outflows, or photoionization from post-AGB stellar populations \citep{Ho1993,Filippenko1996,Bremer2013,Singh2013}. 
LINER galaxies are the largest AGN sub population, and may constitute one-third of all nearby galaxies \citep{Heckman1980,Ho1997e}, yet \bco~ was one of only three cases of a CLAGN in a LINER out of the nearly 70 known CLAGN at the time.\footnote{We note that the other two known so-called CL LINERs, NGC 1097 \citep{Storchi1993} and NGC 3065 \citep{Eracleous2001} are, or are reminiscent of, transient double peaked emitters, which may be distinct from changing-look AGN.} 
Furthermore, as a LINER, \bco~ had a lower inferred accretion rate in its low state \citep[$L/L_{\rm{Edd}} \lesssim 0.005$,][]{Gezari2017} compared to the majority of previously discovered CLAGN \citep{MacLeod2019}, implying a much more dramatic transformation.

We report the discovery of \num~new CLAGN, all classified as LINER galaxies by their archival SDSS spectra, detected as nuclear transients by the Zwicky Transient Facility (ZTF; \citet{Graham2019,Bellm2019a}), and spectroscopically confirmed as ``changing-look'' to a NLS1 or broad-line (type 1) AGN spectral class.  One of these nuclear transients, \tyrion, was initially classified as a candidate tidal disruption event (TDE) from the presence of Balmer and He~II emission lines  \citep{Arcavi2018}. 
Here, we show that the ZTF light curve, together with our sequence of follow-up optical spectra and UV and X-ray monitoring with {\it Swift} and follow-up UV spectra with {\it HST}, are more consistent with a CLAGN classification. 
It was previously thought that, although they are commonly found in Seyferts, coronal emission lines (such as [Fe~VII] $\lambda$6088) should never be exhibited by LINER-like galaxies by definition (e.g. \citealt{Corbett1996}).
 However, here we also report the surprising appearance of coronal lines coincident with an increase in UV/optical and soft X-ray continuum emission and broad Balmer emission consistent with a NLS1 in this galaxy previously classified as a LINER.
 
This paper is organized as follows.  In Section~\ref{sec:obs}, we present our sample selection of nuclear transients in LINERs, information on the host galaxies, ZTF and archival optical light curves, optical spectroscopic observations, and multiwavelength follow-up observations of \tyrion, including details of the data reduction involved.  
In Section~\ref{sec:spc_analysis}, we introduce a new class of changing-look LINERs, and compare their properties to previously reported Seyfert CLAGN, focusing on the particularly interesting case of \tyrion, which transformed from a LINER to a NLS1. In Section~\ref{sec:disc} we discuss the results of our analysis, the conclusions of which are summarized in Section~\ref{sec:concl}.

Throughout the paper we use UT dates, and assume the following cosmology for luminosity calculations: $H_0$ = 70 km s$^{-1}$ Mpc$^{-1}$,  $\Omega_\Lambda$= 0.73 and $\Omega_M$ = 0.27. We have corrected for Galactic extinction toward the sources where explicitly stated. 
All magnitudes are in the AB system, and all uncertainties are at the 1$\sigma$ level unless otherwise noted. We adopt the definition\footnote{{Our sample is not limited to these magnitudes, this criterion is merely used to distinguish quasars from Seyfert AGN.}} for a quasar from the SDSS DR7 quasar catalog \citep{Schneider2010} as having an absolute $i$-band magnitude brighter than $-22$.

\section{Discovery and Observations} \label{sec:obs}
\subsection{Sample Selection Criteria}
\label{sec:select}
We selected CLAGN candidates first flagged as nuclear transients in the ZTF alert stream (described further in Section~\ref{sec:ztflc}) and with a cross-match within 1\farcs0~of a LINER or type 2 Seyfert galaxy in the Portsmouth Catalog's narrow-line ratio BPT classifications\footnote{\url{https://www.sdss.org/dr12/spectro/galaxy_portsmouth/\#kinematics}} \citep{Bolzonella2000, Thomas2013}. Those  classifications, described further in Section~\ref{sec:liner}, are based on stellar population and emission line fits to SDSS DR12 spectra, performed with Penalized Pixel Fitting (\texttt{pPXF}; \citet{Cappellari2004,Cappellari2016}) and Gas and Absorption Line Fitting (\texttt{GANDALF}; \citet{Sarzi2017}), respectively. In this study, we focus on the ``LINER CLAGN'' that emerged as a new class of changing-look AGN and display the most dramatic spectral variability of the CLAGN in our ZTF sample (we reserve discussion of the complementary sample of Seyfert CLAGN for a forthcoming publication).  

\subsection{ZTF Light Curve} \label{sec:ztflc}
ZTF surveys the extragalactic\footnote{Additional public and private allocations are made to survey the Galactic Plane at higher cadence. See \citet{Bellm2019b} for details.} Northern Sky in two modes:  a public Mid-Scale Innovations Program (MSIP) survey of 15,000 deg$^{2}$ of sky every 3 nights in $g$ and $r$ filters, 
and a high-cadence ZTF partnership survey of 3400 deg$^{2}$ with a dense cadence of 6 epochs each in $g$ and $r$ filters per night. It also surveys in $i$-band every 4 nights with a footprint of 10725 deg$^2$ \citep{Bellm2019b}.  
PTF and iPTF (2009$-$2016; \citet{Law2009,Rau2009}) also utilized Palomar Observatory's Samuel Oschin 48" Schmidt telescope; the camera upgrade for ZTF has a 47 deg$^2$ FoV and reaches 20.5 $r$-band mag in 30 seconds exposures, with a  more efficient areal survey speed  of 3760 deg$^{2}$ hr$^{-1}$. 
Images are processed each night by the Infrared Processing and Analysis Center (IPAC) pipeline \citep{Masci2019}, where difference imaging and source detection are performed to produce a transient alert stream \citep{Patterson2019}, distributed to the GROWTH Marshal \citep{Kasliwal2019} and other brokers via the University of Washington Kafka system.  \citet{VanVelzen2019} presented details of the nuclear transients filtering procedure.

All transients in the sample were discovered in 2018 between April and November, 
all in the ZTF MSIP survey (specific dates are summarized in Table~\ref{tab:basic}).  ZTF18aajupnt\footnote{\label{foot:GOT}
As ZTF given names are typically a mouthful of letters (appropriately so, due to the requirement of naming upwards of a million alerts per night), the ZTF Black Holes Working Group has informally begun naming TDEs from a fictional world with no shortage  of characters: HBO's {\it Game of Thrones}. As it was initially thought to be a TDE, \tyrion~was affectionately dubbed ``Tyrion Lannister''.} 
~was also detected in the ZTF Partnership survey on 2018 May 31, and (as it was detected in both surveys in the same night) was registered publicly to the Transient Name Server (TNS) as AT2018dyk. 
Transients were required to have a real-bogus (RB) score $\geq 0.5$ as classified by ZTF machine learning \citep{Mahabal2019}. Further details on the transients, including discovery difference absolute magnitudes, are in Table~\ref{tab:basic}. 

\setlength{\tabcolsep}{2pt}
\begin{table*}[ht!]

\caption{Basic data for the changing-look LINER sample. We list redshifts from the Portsmouth SDSS DR12 catalog \citep{Thomas2013}, which is described in Section~\ref{sec:select}. 
Transition timescales $\delta t$ {(observer frame)} are roughly constrained based on the time delay between the onset of variability detected in the host in the archival light curves, and the time of the first spectrum taken in the type 1 AGN state. 
Estimates of star formation rate by \citet{Chang2015} are from SDSS+WISE SED model fitting. $\Delta m$ is the variability magnitude change defined in Eq. 3 of \citet{Hung2018} as $\Delta m=-2.5\rm{log}(10^{-m_{r,\rm{host}}/2.5} + 10^{-m_r/2.5 })-m_{r,\rm{host}}$, where $m_r$ represents the brightest, transient $r$-band magnitude in the difference-imaging light curve. \tyrion, described further in Section~\ref{sec:tyrion} is the least luminous transient, and has the nearest host of the sample. 
}
\begin{tabular}{llllllllllll}
\label{tab:basic}
Name                   & RA          & Dec         & $z$ & $D_{\rm{Lum}}$  & Discovery{{/Follow-up}}   & $M_{\rm{Discovery}}$  & $\delta t $ & Host\footnote{\citet{Kuminski2016}}      & log SFR & $\Delta m_{\rm{var}}$ & High State                            \\ 
&  (hh:mm:ss.ss)  &  (dd:mm:ss.ss) & & (Mpc) & & (mag) & (yr) & & [$M_\odot$ yr$^{-1}$] & (mag) & \\ \hline 
{{(A)}} ZTF18aajupnt\footnote{hosted in SDSS J153308.02+443208.4/IRAS F15313+4442/2MASX J15330803+4432086 } & 15:33:08.01 & +44:32:08.2 & 0.0367 & 158 & 2018 May 31\footnote{In ZTF Partnership Survey}{{/June 12}}\footnote{{Listed is the first spectroscopic follow up of \tyrion. The full campaign of optical spectroscopic follow-up of this source is summarized in Table~\ref{tab:spc}}} & $-16.59$            & \textless 0.3    & {{SBb D}} &   0.177               & $-0.18$                 & NLS1                             \\
{{(B)}} \noname\footnote{TNS name AT2018cdp; hosted in SDSS J113355.93+670107.0/2MASX J11335602+6701073}  & 	11:33:55.83  & +67:01:08.0 & 0.0397  & 171 & 2018 May 10{{/June 21}} & $-17.80$            & \textless 6.8    & {{SBa(r)}}\footnote{\citet{Hernandez2010} (\citet{Kuminski2016} reported nearly a 50\% probability for both spiral and elliptical type for this host galaxy.)}  & 0.147                 & $-0.06$                 & Seyfert 1                             \\ 
{{(C)}} \bronn\footnote{hosted in SDSS J125403.78+491452.8/2MASX J12540375+4914533
}  & 12:54:03.80 & +49:14:52.9 & 0.0670  & 296 & 2018 April 8{{/Apr 11}} & $-18.25$            & \textless 0.6    & Elliptical     & $-0.058$                & $-0.12$                 & quasar                             \\
{{(D)}} \varys\footnote{TNS name AT2018ivp; hosted in SDSS J091531.04+481407.7}  & 09:15:31.06 & +48:14:08.0 & 0.1005  & 457 & 2018 April 11{{/May 06}} & $-19.09$            & \textless 0.7    & {{Sb D} } & 0.092                 & $-0.29$                 & quasar                             \\ 
{{(E)}} \jorah\footnote{TNS name AT2018gkr; hosted in SDSS J081726.41+101210.1/2MASX J08172642+1012101} & 	08:17:26.42 & +10:12:10.1 & 0.0458 & 199 & 2018 Sept 15{{/Dec 09}}\footnote{{Listed is the first spectroscopic follow up of \jorah. Additional high-resolution spectroscopic follow-up of this source was taken on 2019 May 02 and is shown in Figure~\ref{fig:spc}.}} & $-17.62$ & \textless 2.6 & Elliptical & 0.227 & $-0.81$ & quasar \\
{{(F)}} \podrick\footnote{hosted in SDSS J122550.30+510846.3/2MASX J12255033+5108461} & 	12:25:50.31 & +51:08:46.5 & 0.1680 & 813 & 2018 Nov 1{{/Dec 03}} & $-20.40$ & \textless 5.3 & Elliptical & 1.267 & $-0.72$ & quasar \\
\hline
\end{tabular}
\end{table*}

The optical photometry for \tyrion, \noname, 
\bronn, \varys, \podrick, 
and \jorah~is comprised of 398, 200, 35, 35, 143, and 207 images, respectively, shown in Figure~\ref{fig:lc}. 
We consider only observations with difference image detections classified as real (with RB score$\geq0.5$ on a scale where 0 is bogus and 1 is real). 
The ZTF optical difference imaging light curves show only the transient nuclear emission in the $g$- and $r$-bands. 
The transients are localized to within 0\farcs19$^{+0{\tiny\farcs}28}_{-0{\tiny\farcs}19}$ (\tyrion), 0\farcs09$\pm$0\farcs26 (\noname),  0\farcs11$^{+0{\tiny\farcs}33}_{-0{\tiny\farcs}11}$ (\bronn), 0\farcs06$^{+0{\tiny\farcs}33}_{-0{\tiny\farcs}06}$ (\varys), 0\farcs10$^{+0{\tiny\farcs}20}_{-0{\tiny\farcs}10}$ (\podrick), and 0\farcs15$\pm$0\farcs15 (\jorah) of their host galaxy nuclei, well within our nuclear selection criterion of $< 0\farcs5$.

\def\names{{ZTF18aajupnt_lc_pd/\tyrion},{ZTF18aasuray_lc_pd/\noname},{ZTF18aahiqfi_lc_pd/\bronn},{ZTF18aaidlyq_lc_pd/\varys},{ZTF18aaabltn_lc_pd/\jorah},
{ZTF18aasszwr_lc_pd/\podrick}}

\begin{figure*}[!htbp]
\centering
\foreach \name/\subcap in \names{%
\subfigure[\subcap]{
  \includegraphics[width=0.4\textwidth]{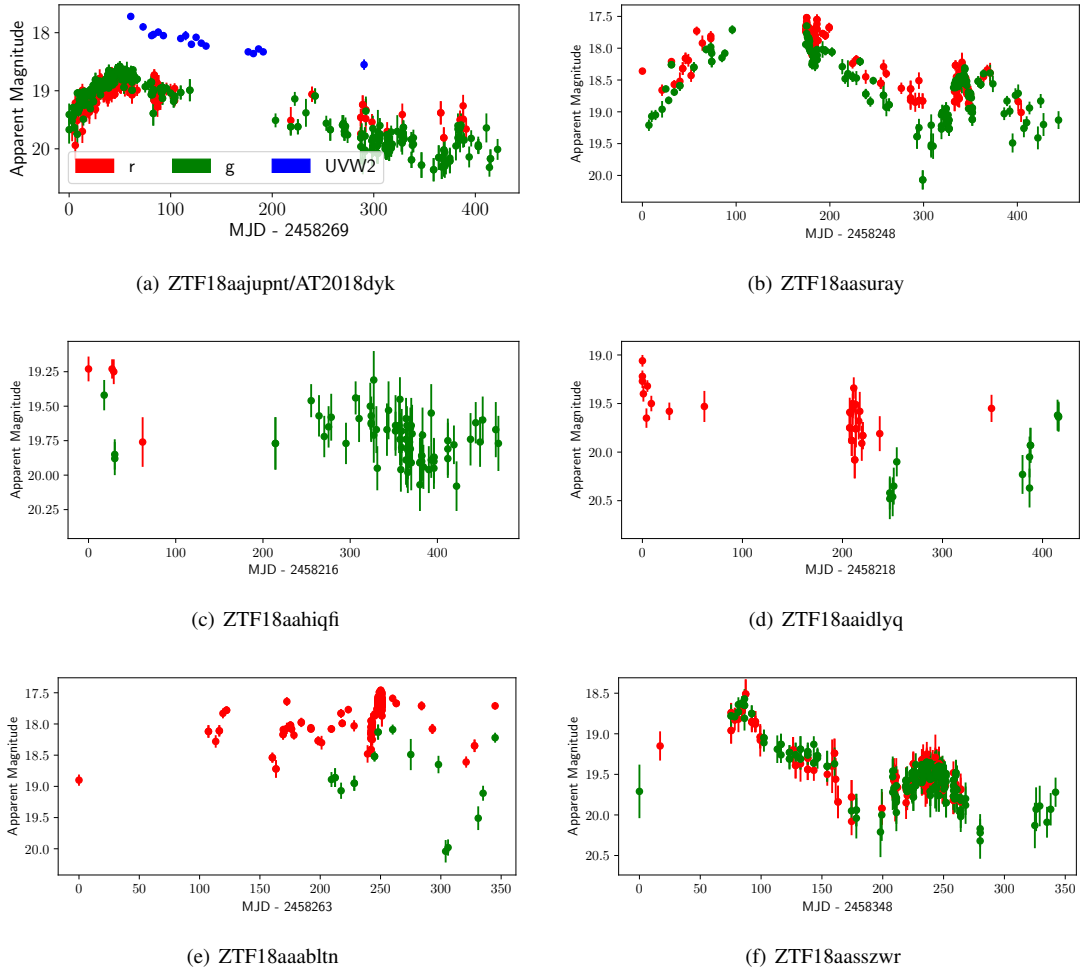} 
  }
  }
\caption{Light curves of the CL LINER sample. Red points represent $r$-band difference imaging photometry data taken with the Palomar 48-inch (P48), green points $g$-band difference imaging photometry, and the blue points are the UVW2 {\it Swift} photometry in the light curve of \tyrion, which tracks the plateau in the optical uncharacteristic of either TDEs or SNe. Note the differences in scale. 
}
\label{fig:lc}
\end{figure*}%

To quantify the amplitude of the flux increase relative to the host galaxy flux, and to compare to variability of CLAGN measured from imaging surveys that do not perform image subtraction, as in \citet{Hung2018}, we add the flux of the host galaxy to the transient flux, to get a variability amplitude, $\Delta m_{\rm{var}} = m_{r,{\rm tot}} - m_{r,\rm {host}}$, where {$m_{r,{\rm tot}} = -2.5\log(10^{-m_{r}/2.5}+10^{ -m_{r,\rm{host}}/2.5})$}, $m_{r}$ represents the brightest transient ZTF $r$-band magnitude{,} and $m_{r,\rm{host}}$ is the archival host magnitude from SDSS DR14. We find $\Delta m_{\rm{var}}$ values ranging from $-0.12$ to $-0.81$ mag for the sources in our sample, with 3 out of 5 below the CLAGN candidate selection criteria of an amplitude of $\Delta r > 0.5$ mag between SDSS and Pan-STARRS1 imaging observations adopted by \citet{MacLeod2019}. 

ZTF18aajupnt (AT2018dyk; discussed more in Section~\ref{sec:tyrion}), \noname, and \podrick~display a slow  months-long rise and plateau (although a visibility gap makes this  unclear for \noname) with a constant color, and gradual decline, with \podrick~exhibiting a second rise and \tyrion~growing redder in the latest observations. All other transients in the sample show flaring in the light curves (see Figure~\ref{fig:lc}) but with less distinct trends,  characteristic of broad-line AGN variability viewed in difference imaging \citep{Choi2014}.

\subsection{Capturing the Transition in Archival Light Curves }  \label{sec:archival}

\begin{figure*}
\includegraphics[width=1.05\textwidth]{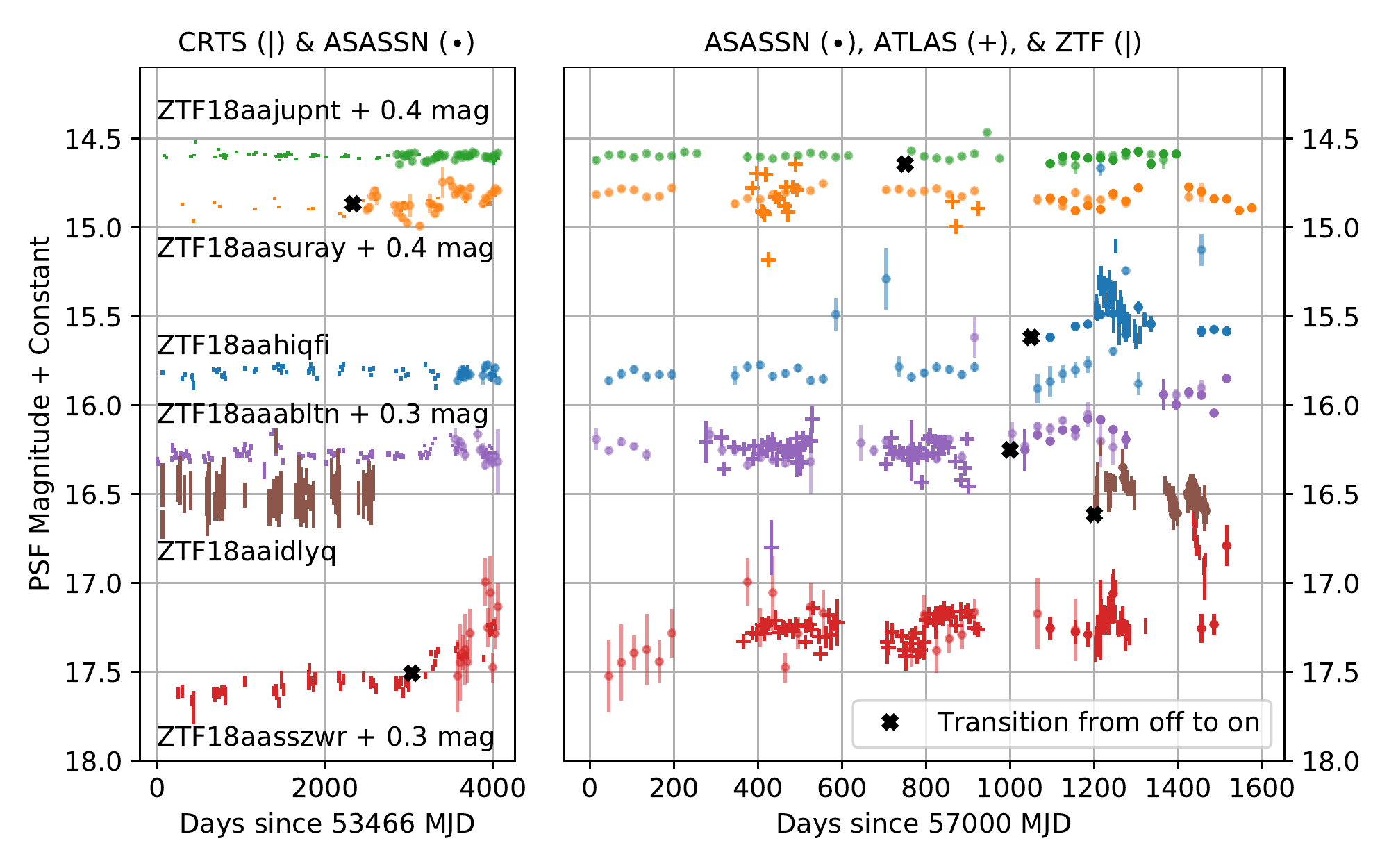}
\caption{Archival light curves of the CL LINER sample summarized in Section~\ref{sec:archival}. The left panel shows years to decades of quiescence (in the ``off'' state while these were still LINER galaxies) observed by CRTS, followed by slow flares in the faintest sources \podrick, \jorah,~and \bronn. The right panel shows the rise, flaring, and decline of the sources caught by ZTF+ATLAS+ASASSN $g$-band observations at these various stages. 
The estimated transition time listed in Section~\ref{sec:archival} for each object is marked by a black ``$\times$''. {This was determined by inferring by-eye approximately when in time (observer frame) the onset of prolonged optical variability from quiescence took place in each source.} When two filters are shown for the same instrument, the redder is shown as more transparent, as in the case of the ASAS-SN $g$ and $V$ photometric points shown. }

\label{fig:archival}
\end{figure*}

Although difference imaging is a useful real-time discovery mechanism for these nuclear transients, archival optical photometric observations can fill in the details of the timing of the transition to its "on" state. With archival light curves extending over a baseline of 13 years from the Catalina Real-time Transient Survey \citep[CRTS;][]{Drake2009}, the All-Sky Automated Survey for Supernovae \citep[ASAS-SN;][]{Shappee2014,Kochanek2017}\footnote{\url{http://www.astronomy.ohio-state.edu/\textasciitilde assassin}}, and Asteroid Terrestrial-impact Last Alert System \citep[ATLAS;][]{Tonry2018}, and ZTF aperture photometry from the IPAC pipeline measured from the static images, we uncover an intriguing uniformity in the events  (Figure~\ref{fig:archival}). Each source in the sample went from lacking any significant variability to flaring dramatically and, for those observed long enough, subsequently declining (\jorah~continues to rise smoothly). 
As not all sources in the sample have peaked, we define the transition timescale for each source reported in Table~\ref{tab:basic} as being from the onset of each flare to the spectroscopic confirmation of the appearance of a blue continuum and broad line emission (except \bco, for which the onset time was constrained by archival and follow up X-ray observations; \citet{Gezari2017}). Turn-on timescales, absolute $r$-band magnitudes at the time of detection with ZTF, variability amplitude relative to the host galaxy flux, and new AGN class following the change are summarized in Table~\ref{tab:basic}  for all transients in the sample. We discuss the details of each source's flaring individually below. 

{\it {(A)} \tyrion} \textemdash~ZTF-matched aperture photometry in $g$ band shows that \tyrion~began flaring some time before 2018 March (58200 MJD) $\sim$2 months prior to discovery in difference imaging on 2018 May 31, and 3 months prior to confirmation of a spectroscopic change. The most recent difference imaging photometry shows a slow decline at constant color. 
Transition timescale: $<0.3$ years, the fastest in the sample.

{\it {(B)} \noname} \textemdash~Discovery with ZTF difference imaging occurred on 2018 May 10 and shows a slow symmetric rise and decline lasting 300 days. \noname~displayed flaring in ASAS-SN data beginning around 2011 Aug (55800 MJD), 6.8  years prior to spectroscopic confirmation of the changing look which occurred on 2018 June 21. Prior to this flaring, Catalina Real-time Transient Survey (CRTS shown in the left panel of Figure~\ref{fig:archival}; \citet{Drake2009}) observations in $V$-band showed no variability above the 0.1 mag level. Transition timescale: $<6.8$ years.

{\it {(C)} \bronn} \textemdash~The rise (seen in ZTF $g$-band matched photometry) starts approximately at 2017 Sept (58000 MJD), 7 months prior to its spectroscopic change. It peaks around 2018 May (58250 MJD; $\sim$1 month after discovery with ZTF difference imaging on 2018 April 8) and subsequently shows a sharp decline. Prior to this flaring, Catalina Real-time Transient Survey (CRTS shown in the left panel of Figure~\ref{fig:archival}; \citet{Drake2009}) observations in $V$-band and ASAS-SN showed no variability above the 0.1 mag level. Transition timescale: $<0.6$ years.

{\it {(D)} \varys}  \textemdash~This source displayed a slight flare in ASAS-SN data just after 2017 Sept (58000 MJD), 7 months prior to detection in ZTF difference imaging and 8 months prior to spectroscopic confirmation of the existence of a BLR, but was faint and quiescent in CRTS beginning in 2005 May (note that this source is near a bright star). Transition timescale: $<0.7$ years.

{\it {(E)} \jorah} \textemdash~CRTS, ATLAS and ASAS-SN show a continuous rise starting around 2016 April (57500 MJD) but this disregards some slight flaring (by 0.2 mag) events at 2008 Nov and just before 2014 Dec (57000 MJD), with both returning to very flat pre-activity levels. This constrains the spectroscopic change to happening within 1000 days ($<2.7$ years) of the flare start time, the first large flare occurring within 9 months of being observed to be a LINER in 2007 Feb. Transition timescale: $<2.6$ years.

{\it {(F)} \podrick} \textemdash~The rise is visible in CRTS around 2018 July (56500 MJD), after which it may have plateaued for a time. Most recently there has been a sharp rise and decline around 2018 May (58250 MJD), with the peak reaching $>1$ mag above original levels. The transition from quiescence thus happened roughly in real time, and was observed with difference imaging 4 months after the flaring began, with the spectroscopic change confirmed within 5.3 years of the initial rise time, and within 5 months of the onset of the most recent flare. We note that two decades ago, \podrick~was a variable ($\rm{rms}=0.14$ mJy) radio source between the NRAO VLA Sky Survey  and Faint Images of the Radio Sky at Twenty centimeters (NVSS and FIRST;  \citealt{Ofek2011}), with a peak flux density at 1.4 GHz of $F_\nu=2.17$ mJy. 
Transition timescale: $<5.3$ years.

{\it \bco} \textemdash~CRTS photometry shows a flare beginning around 2012 March (56000 MJD), 8 years after being observed to be a LINER and 4 years prior to discovery and classification of a quasar in iPTF, and the latest ZTF $g$-band data show it declining rapidly.  However, archival XMM Slew Survey observations constrain the onset of the X-ray source detected by Swift in its broad-line state to $<1.1$ years before \citep{Gezari2017}.  
Transition timescale: $<1.1$ years.

\subsection{Host Galaxy Morphology}
Images of the \num~transients' host galaxies from SDSS are shown in Figure~\ref{fig:cutouts}, and basic data including the hosts' names, matched coordinates, redshifts, luminosity distances, morphological types, and star formation rates (SFRs) are summarized in Table~\ref{tab:basic}. The SFR estimates by \citet{Chang2015} were obtained through Multi-wavelength Analysis of Galaxy Physical Properties (MAGPHYS; \citet{daCunha2012}) model fitting of dust extinction/emission, and SEDs constructed from WISE+SDSS (WISE: \citet{Wright2010}) matched photometry of present-epoch galaxies (we note that SFRs for only two AGN in our sample were measured by \citet{Chang2015}, the rest did not fit their criteria). 
The bulges of the LINERs' hosts are similar in apparent color and extent, but the host of \varys~exhibits evidence for a bar and ring, and the host of~\jorah~exhibits apparent elongation. The host of \tyrion~stands out in the sample as the only gas-rich spiral galaxy, and we note that NLS1s typically occur in spiral-type galaxies \citep{Crenshaw2003}. 
Black hole masses estimated from the host galaxy luminosity, bulge mass, and velocity dispersions derived from the SDSS host imaging and spectra have been measured in Section~\ref{sec:mbh} and are summarized in Table~\ref{tab:mbh}.

\def\names{{tyrion_sdss_cutout/\tyrion},
{ZTF18aahiqfi_sdss_cutout/\bronn},
{ZTF18aaidlyq_sdss_cutout/\varys},
{\noname_sdss_cutout/\noname},
{ZTF18aasszwr_sdss_cutout/\podrick},
{ZTF18aaabltn_sdss_cutout/\jorah}}

 \begin{figure}[htbp]
\begin{center}
\foreach \name/\subcap in \names {
        \subfigure[\subcap]{
        \includegraphics[width=0.2\textwidth]{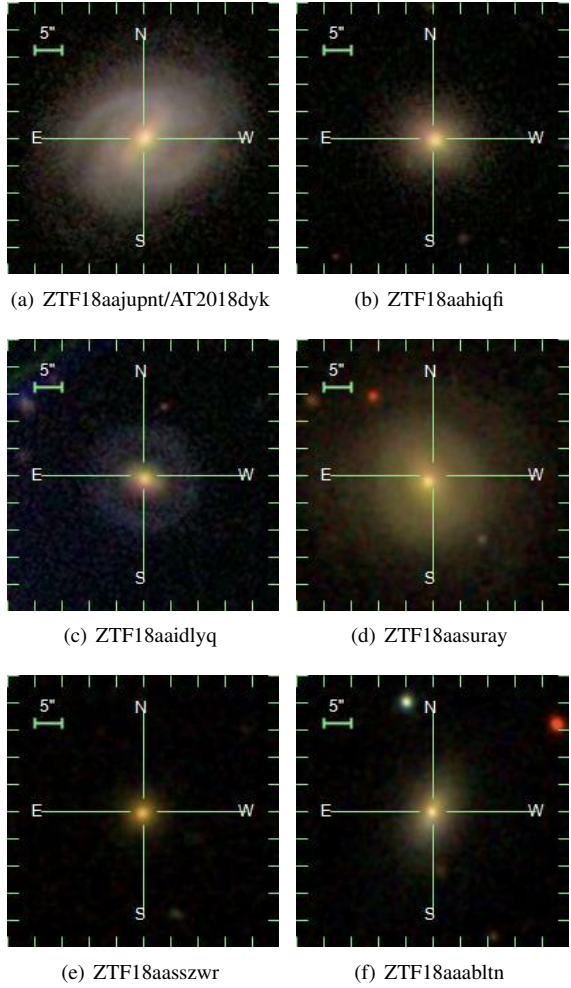}}
}
\end{center}
\caption{Composite $ugriz$ color SDSS images of the host galaxies of the changing-look LINER sample. Their individual morphological classifications are listed in Table~\ref{tab:basic}.}\label{fig:cutouts}
\end{figure}

\setlength{\tabcolsep}{3pt}
\begin{table*}[ht!] 
\caption{
Properties of the host galaxies of our sample of changing-look LINERs from ZTF and iPTF.    
We also show $M_{\rm{BH}}$ calculated in Section~\ref{sec:mbh} from the host galaxy luminosity, mass, and velocity dispersion, respectively.
}
\begin{tabular}{llllcllcccl}
\label{tab:mbh}

Name                   & $M_{r,\rm{host}}$\footnote{Computed from the $r$-band de Vaucouleurs / exponential disk profile model fit magnitude from the SDSS DR14 photometric catalog.} & log $M_{\rm{Bulge}}$\footnote{Computed from broadband SED fits to photometric measurements of SDSS DR 7 galaxies \citep{Mendel2014}.} & $\sigma_\bigstar$\footnote{Measured from the SDSS spectrum using the \texttt{PPXF} method.} & $\lambda L_{5100A}$  & FWHM$_{H_{\beta}}$ & log $M_{\rm{BH},M_r}$\footnote{\citet{McLure2002}} & log $M_{\rm{BH,Bulge}}$\footnote{\citet{Haring2004}} & log $M_{\rm{BH},\sigma\bigstar}$\footnote{\citet{Tremaine2002}} & log $M_{\rm{BH,vir}}$  & $L/L_{\rm{Edd}}$\footnote{In high state; $M_{\rm{BH},\sigma\bigstar}$ was employed to obtain the black hole masses used in computing the Eddington
ratio (see Section~\ref{sec:ledd}).} \\ %
& (mag) & [$M_\odot$] & (km s$^{-1}$) & ($10^{43}$ erg s$^{-1}$) & (km s$^{-1}$) & [$M_\odot$] & [$M_\odot$] & [$M_\odot$] & [$M_\odot$] & \\ \hline

ZTF18aajupnt & $-22.00$            & 10.66$\pm$0.15               & 150               &  0.23$\pm$0.02                                      & 939$\pm$28         & 8.0                   & 7.8                     & 7.6                              &  5.5                  &  0.004                      \\
\noname & $-21.70  $          & 10.73$\pm$0.15               & 230               & 0.62$\pm$0.05                                  & 4270$\pm$218         & 7.9                   & 7.9                     & 8.4                              & 7.1                   &  0.002                     \\ 
ZTF18aaidlyq           & $-21.64$            & -                    & 120               & 1.15$\pm$0.04                                      & 7726$\pm$458       & 7.9                   & -                       & 8.2                              & 7.8                   & 0.005                    \\
ZTF18aahiqfi           & $-21.63$            & -                    & 210               & 0.40$\pm$0.01                                       & 8809$\pm$723       & 7.9                   & -                       & 7.2                              & 7.6                   &  0.02                     \\
\podrick & $-22.19$ & 11.19$\pm$0.15 & 180 & 5.7$\pm$0.3  & 6461$\pm$846 & 8.1 & 8.3 & 7.9 &  8.1 & 0.05\\
\jorah & $-20.62$ & - & 140 &  0.56$\pm$0.05 & 3057$\pm$648 & 7.3 & - & 7.5 &  6.8 &  0.01 \\
\bco             & $-22.21 $           & -                    & 176               & 6.9$\pm$0.2                                         & 4183$\pm$213       & 8.4                   & -                       & 7.9                              &  7.8                   & 0.06                      \\\hline  
\end{tabular}
\end{table*}

\subsection{Optical Spectroscopy}
\label{sec:spc}
We obtained spectral follow-up of nuclear transients in known LINERS and Sy~2 galaxies as described in Section~\ref{sec:select} to confirm changing-look AGN candidates, as neither ``true'' narrow-line Sy~2s nor LINERs are expected to vary significantly.\footnote{{Curiously, long-term X-ray (e.g. \citealt{ HernandezGarcia2013}) and compact nuclear UV \citep{Maoz2005} variability by a factor of a few has been observed in a number of both broad and narrow type LINERs, attributed to an advection dominated accretion flow mechanism in an AGN component in the former work and a ``scaled-down'' Seyfert analog in the latter.}}

We observed \bronn, \varys, and \noname~with the Deveny spectrograph on the Discovery Channel Telescope (DCT; spectral coverage of 3600-8000 \AA)  with a 1\farcs5 wide slit, central wavelength of 5800 \AA~and exposure times of $2 \times 900$, $2 \times 1200$, and 1400 seconds on 2018 April 11, May 06, and June 21, respectively. 
The DCT spectra were reduced with standard IRAF routines, corrected for bias and flat-fielding, and combined into a single 2D science frame. Wavelength and flux calibration were done via a comparison with spectra of an arc lamp and the flux standard Feige 34, respectively. The spectra have not been corrected for telluric absorption. We found that the Balmer lines of \bronn, \varys, and \noname~had gotten dramatically stronger and broader compared to archival SDSS spectra of their hosts, obtained more than a decade prior (in April 2003, Dec 2002, and Feb 2001, respectively). 

\podrick~and \jorah~showed similar striking spectral changes when they were followed up on 2018 Dec 3 and 9 using the Spectral Energy Distribution Machine (SEDM; \citealt{Blagorodnova2018}) IFU spectrograph on the Palomar 60-inch (P60; \citealt{Cenko2006}) operating as part of ZTF. Both displayed broader emission lines and bluer continuua compared to archival LINER spectra (from Feb 2007 and Jun 2004, respectively). 
The SEDM data were reduced with \texttt{pySEDM} \citep{Rigault2019}. 
{\jorah~was later followed up with DCT on May 02 2019.}

See the spectral comparisons for all CLAGN in the sample in Figure~\ref{fig:spc}, and zoom-ins of the emission lines in the ``off'' states in Figures~18, 19, and ``on'' states in Figure~20 of the Appendix (available in the electronic version).  
The hosts of all \num~transients in this sample were originally classified as LINERs in SDSS, however we re-measured the diagnostic narrow-line ratios in Section~\ref{sec:liner}, and find that the majority of the sample is on the borderline between a LINER and Seyfert classification. 

\def\names{{16bco_spc/\bco},{ZTF18aahiqfi_spc/\bronn},{ZTF18aaidlyq_spc/\varys},{\noname_spc/\noname},{ZTF18aasszwr_spc/\podrick},{ZTF18aaabltn_spc/\jorah}}

\begin{figure*}[htbp]
\begin{center}
\foreach \name/\subcap in \names {%
        \subfigure[\subcap]{
        \includegraphics[width=0.4\textwidth]{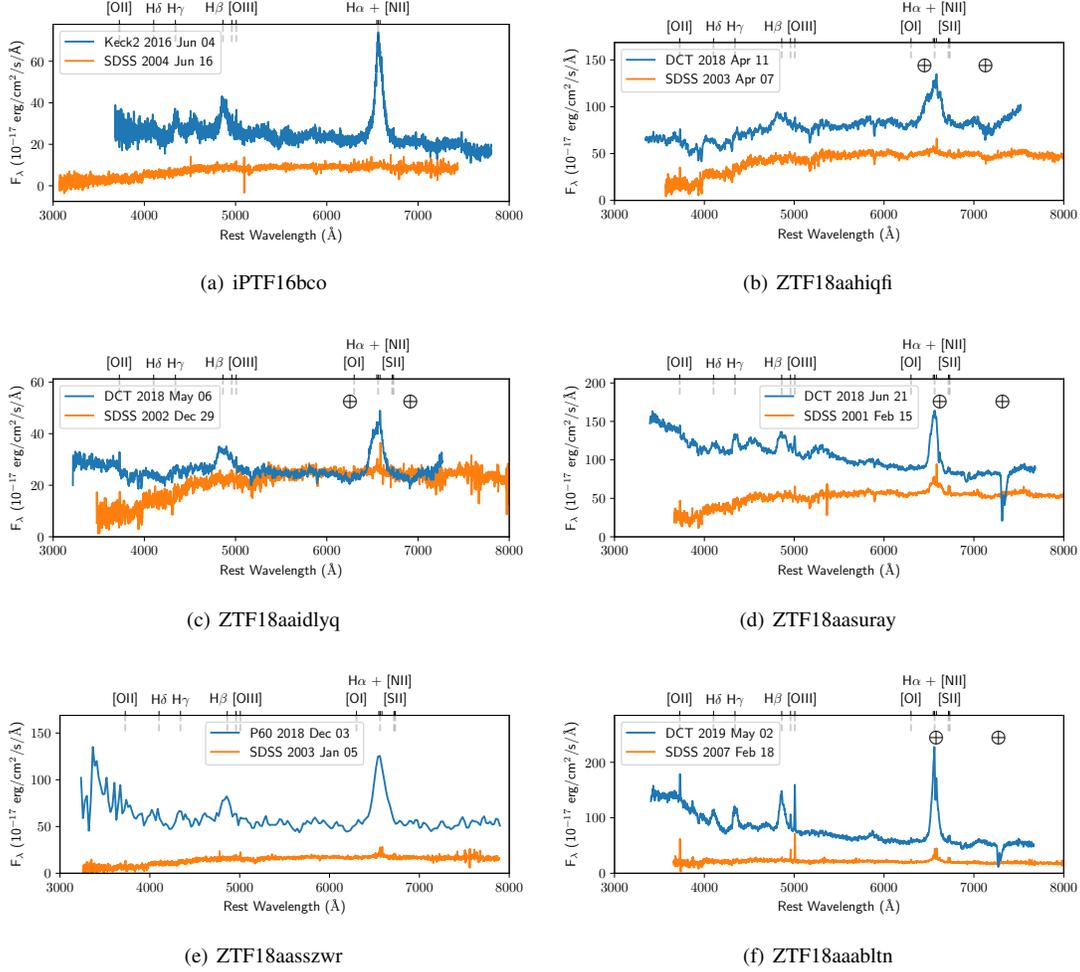}}
}
\end{center}
\caption{Comparison of early and follow-up spectra of the other  CLAGN in the sample. Note that the Palomar 60-inch ``P60'' spectrum has a difference in aperture affecting the flux measurement by a factor of order unity. {The $\bigoplus$ symbols indicate atmospheric telluric absorption bands.} Detailed follow-up of \tyrion~(not shown here) is presented in Figure~\ref{fig:tyrion_spc}.}\label{fig:spc}
\end{figure*}

\begin{figure*}[ht!]
\includegraphics[width=0.9\textwidth]{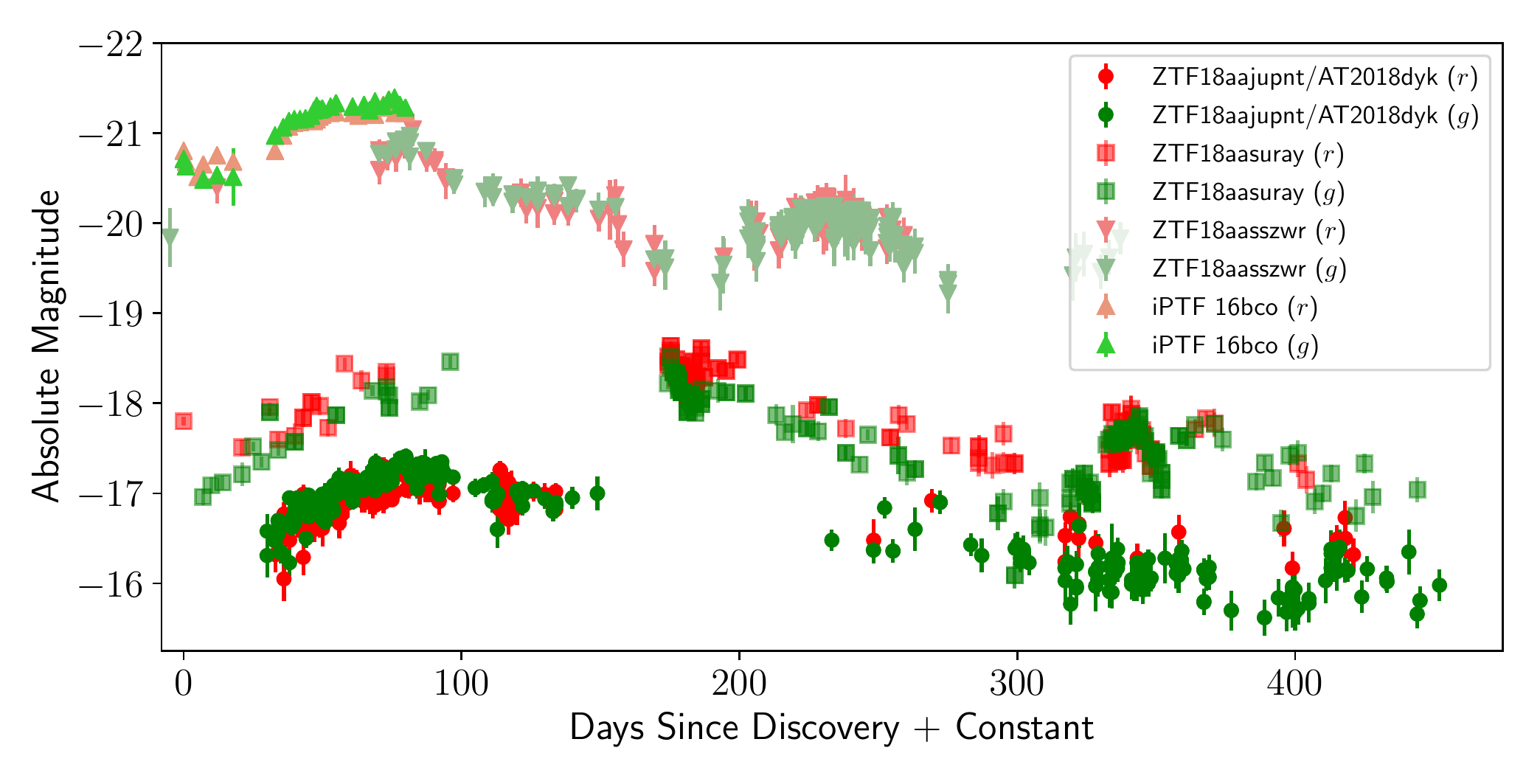}
\caption{Difference imaging light curves of the CL LINERs with the best-sampled P48 observations in the ZTF sample (\tyrion, \podrick, and \noname) plotted in absolute magnitude compared to that of CL LINER \bco~(triangle shaped points). {Red and green colors represent $r$- and $g$-band observations, respectively, with slightly different shades used only to distinguish the different sources.}  \podrick~and \bco~are similar in luminosity and more luminous than \tyrion~and \noname~by about 2.5 mag. \noname~has a much slower evolution and is constantly redder in color, whereas \tyrion~reddens $\sim$280 days into its evolution. The rise of \tyrion~mirrors that of \bco, whereas the decline appears slower than but similar in shape to that of \podrick. 
}
\label{fig:16bco}
\end{figure*}

Due to its similarity to a TDE at early times, we promptly initiated a multi-wavelength follow-up campaign of \tyrion~which we describe in the following sections. 
Following the discovery of a blue continuum with the Double Spectrograph (DBSP) of the Palomar 200-inch Hale telescope on 2018 June 12 (PI: David Cook), we monitored \tyrion~with five additional epochs of optical spectroscopy with SEDM on Palomar's 60-inch on 2018 July 22 and Aug 12, LRIS on the Keck I telescope on 2018 Aug 7 (PI: Kulkarni), Gemini GMOS-N on 2018 Aug 21 (PI: Hung), and with Deveny on the DCT on 2018 Sept 12 (PI: Gezari).  We detail the configurations of the spectroscopic follow-up observations of \tyrion~in Table~\ref{tab:spc}. 
During this time, its optical light curve plateaued in a manner strikingly similar to \bco~ (shown in Figure~\ref{fig:16bco}). It also surprisingly displayed coronal emission lines (those detected are shown in  Figures~\ref{fig:resid} and 21) in a heretofore low-ionization nuclear source.  

\begin{table}[]
\caption{Spectroscopic Legacy and Follow-up Observations of \tyrion. 
}
\label{tab:spc}
\begin{tabular}{@{}llll@{}}
\toprule
Obs UT       & Instrument              & Exposure (s) & Reference                                \\ \midrule
2002 July 11 & SDSS                    & 28816        & \citet{Abolfathi2018} \\
2018 June 12 & Palomar 200" DBSP       & 2400            & This work                                \\ 
2018 July 22 & Palomar 60" SEDM & 2430         & This work                                \\
2018 July 30 & {\it Swift} XRT               & 40400        & This work                                \\
2018 Aug 7   & Keck LRIS             & 300          & This work                                \\
2018 Aug 11  & {\it XMM} EPIC pn             & 11906        & This work                                \\
2018 Aug 12  & Palomar 60" SEDM & 2430         & This work                                \\
2018 Aug 12  & FTN FLOYDS-N            & 3600         & \citet{Arcavi2018}    \\
2018 Aug 21  & Gemini GMOS-N           & 600          & This work                                \\
2018 Sept 1  & {\it HST} STIS                & 2859            & This work                                \\ 
2018 Sept 12 & DCT Deveny              & 2400         & This work                                \\\bottomrule
\end{tabular}
\end{table}

\begin{figure*}[ht!]
\centering 
\includegraphics[width=0.75\textwidth]{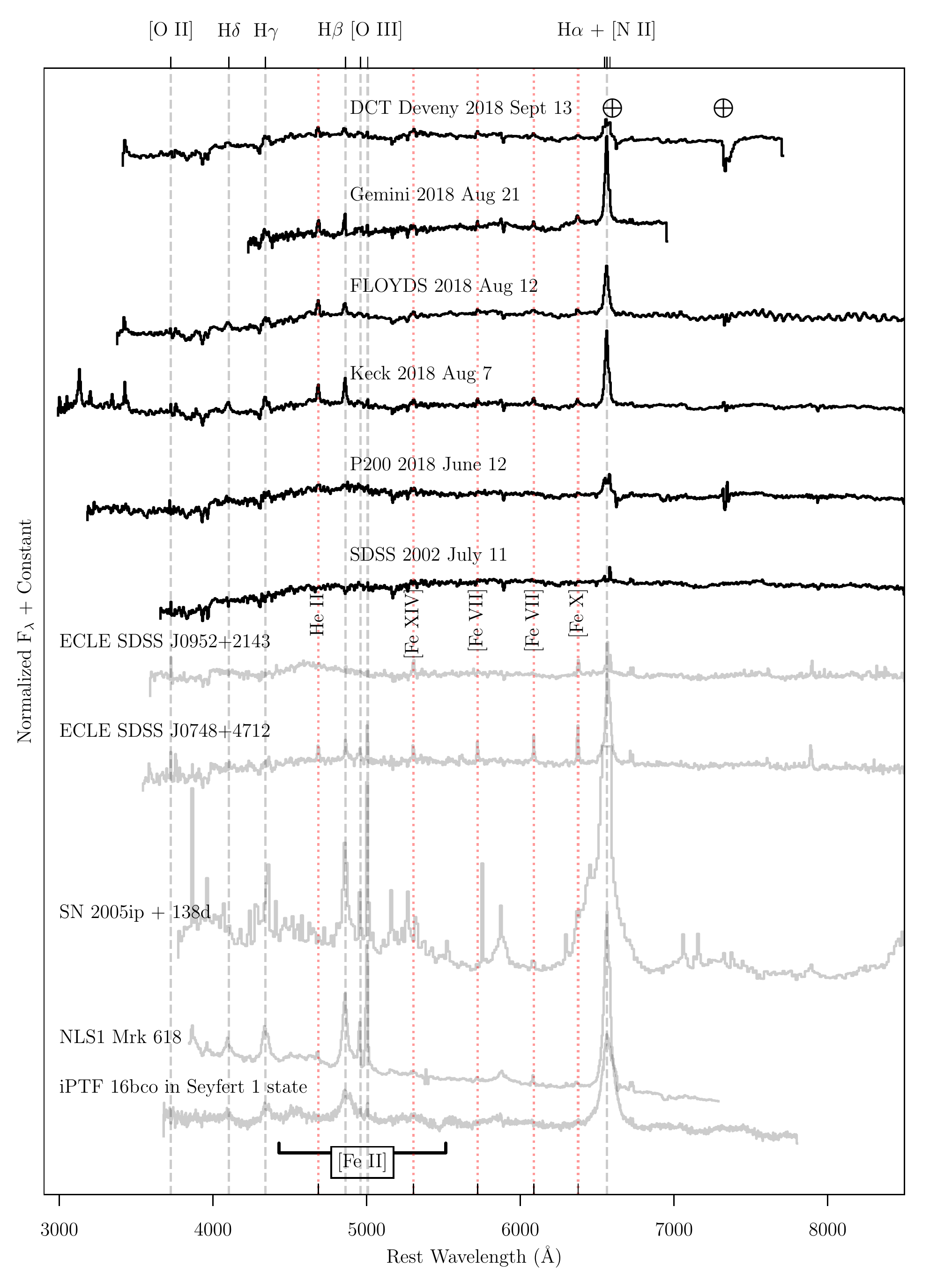}
\caption{Host and follow-up spectra of \tyrion, alongside various AGN and coronal line emitters for comparison. AGN emission lines are annotated in gray and are labeled above the figure. Coronal lines are annotated in red and are labeled in the middle of the figure. {The $\bigoplus$ symbols indicate atmospheric Telluric absorption bands.} The flux of the H$\alpha$ line (only) in SN 2005ip has been truncated for visual purposes (as it lies well above the upper boundary of the plot). Spectra have been rebinned by a factor of four for visual purposes.}
\label{fig:tyrion_spc}
\end{figure*}

\begin{figure*}[ht!]
\includegraphics[width=0.9\textwidth]{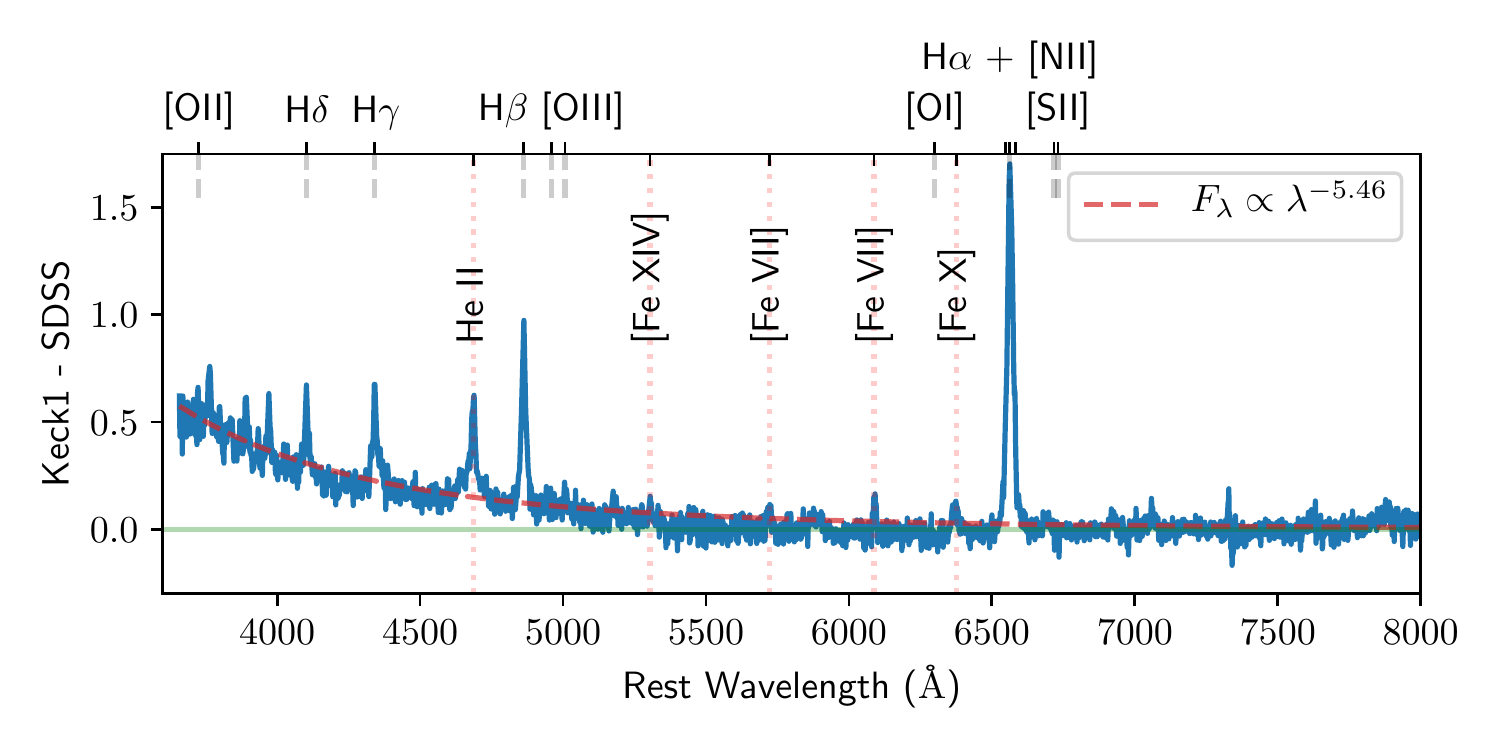}
\caption{Host-galaxy-subtracted Keck1 spectrum of \tyrion~showing presence of coronal emission lines (red dotted lines) and a blue excess in the residuals. 
We fit the non-stellar blue continuum with a power law (red dashed line with $\alpha=-5.46$).}\label{fig:resid}
\end{figure*}

Figure~\ref{fig:tyrion_spc} shows a complete series of spectra obtained for \tyrion, as well as comparisons to some examples of other AGN and transient types, including the class of extreme coronal line emitters (ECLEs) and the luminous SN~IIn SN 2005ip which demonstrated strong coronal line emission \citep{Smith2009}. These spectra were reduced with standard pipelines and procedures for each instrument. Measurements of the flux, luminosity, radial velocity, full-width-at-half-maximum (FWHM), and equivalent width of the emission lines, including the coronal emission lines ([Fe~XIV] $\lambda 5304$, [Fe~VII] $\lambda\lambda 5721, 6088$, [Fe~X] $\lambda 6376$ in the spectrum with the highest signal-to-noise detection of the coronal lines is given in Table~\ref{tab:lines}. The FLOYDS-N spectrum from 2018 Aug 12 was reported by \citet{Arcavi2018} to have broad H$\alpha$, and both broad and narrow H$\beta$ and He~II. At that time, a blue continuum was not obvious in their spectrum. However, we show a power-law blue excess is clearly detected in the residuals of the spectra after subtracting a model for the host galaxy light (Figure~\ref{fig:resid}). 

\begin{table*}[ht!]
\begin{center}
\caption{Line measurements for \tyrion~from fits in Figure~21 (available in the Appendix of the electronic version) and used in Figures~\ref{fig:fwhm}, \ref{fig:bpt} and \ref{fig:haoiii}. The blueshift measured significantly only in Fe~X translates to $\approx$0.0005 c.}
\label{tab:lines}
\begin{tabular}{lcccccc}
\toprule
                        & $\lambda$ (\AA) & $F_{\lambda}$ ($10^{-15}$ ergs s$^{-1}$ cm$^{-2}$ \AA$^{-1}$) & $L$ ($10^{39}$ ergs s$^{-1}$) & $v_r$ (km s$^{-1}$) & FWHM (km s$^{-1}$) &        EW (\AA) \\
\midrule
              H$\alpha$ &       6562.80 &                                     27.67$\pm$0.59 &                  82.4$\pm$1.2 &            $57\pm4$ &        $1061\pm19$ &  $56.9\pm1.5$ \\
   $[\rm{NII}]\lambda$6548 &       6548.05 &                                      0.21$\pm$0.19 &                 0.64$\pm$0.37 &         $-612\pm19$ &         $212\pm59$ &   $0.4\pm0.0$ \\
   $[\rm{NII}]\lambda$6583 &       6583.45 &                                      1.11$\pm$0.15 &                 3.31$\pm$0.29 &          $954\pm10$ &         $335\pm28$ &   $7.9\pm0.2$ \\
               H$\beta$ &       4861.30 &                                      9.02$\pm$0.32 &                26.85$\pm$0.94 &            $76\pm8$ &         $939\pm28$ &  $18.0\pm0.7$ \\
               $[\rm{OIII}]$ &       5006.84 &                                      0.96$\pm$0.16 &                 2.86$\pm$0.47 &           $73\pm24$ &         $489\pm59$ &   $2.1\pm0.3$ \\
                   HeII &       4686.00 &                                      3.48$\pm$0.29 &                10.37$\pm$0.85 &           $10\pm28$ &        $1157\pm69$ &   $6.7\pm0.6$ \\
              $[\rm{FeXIV}]$ &       5304.00 &                                      0.45$\pm$0.14 &                 1.33$\pm$0.40 &           $37\pm44$ &        $546\pm115$ &   $1.0\pm0.3$ \\
 $[\rm{FeVII}]\lambda$5721 &       5721.00 &                                      0.81$\pm$0.14 &                 2.40$\pm$0.41 &           $62\pm40$ &         $795\pm98$ &   $1.6\pm0.3$ \\
 $[\rm{FeVII}]\lambda$6088 &       6088.00 &                                      1.08$\pm$0.13 &                 3.22$\pm$0.39 &           $68\pm22$ &         $600\pm54$ &   $2.3\pm0.3$ \\
                $[\rm{FeX}]$ &       6376.00 &                                      1.83$\pm$0.19 &                 5.44$\pm$0.56 &         $-160\pm36$ &        $1301\pm94$ &   $3.9\pm0.4$ \\
\bottomrule
\end{tabular}
\end{center}
\end{table*}

\label{sec:ext}
We have corrected for Galactic extinction in the spectra in Figure \ref{fig:tyrion_spc}, with color excess $E(B-V)=0.0164$ mag (from the \citet{Schlafly2011} dust map\footnote{\url{https://irsa.ipac.caltech.edu/applications/DUST/}}). We use the optical correction curve for $R_V = 3.1$ given by Eqs. 3.a. and b. in \citet{Cardelli1989}, such that $f_{\rm{corr}} = f_{\rm{obs}}10^{A_\lambda/2.5}$.

\subsection{UV Imaging and Spectroscopy} 
\begin{figure}
\includegraphics[width=0.5\textwidth]{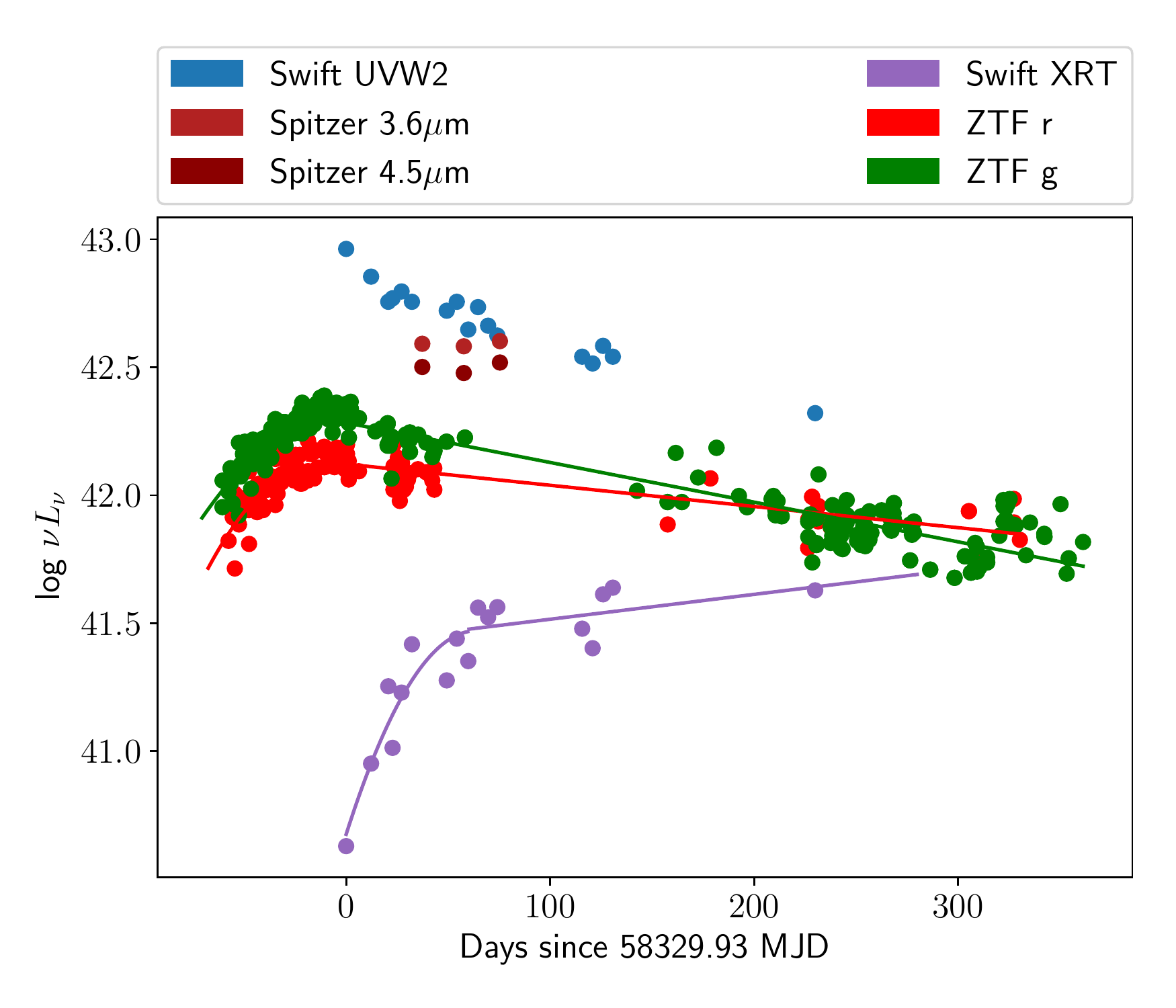}
\caption{The $\nu L_\nu$ light curve of \tyrion,~comparing {\it Spitzer} data to concurrent {\it Swift} UVOT, XRT and ZTF observations. For the {\it Spitzer} and {\it Swift} UVOT observations we subtracted the host galaxy light as estimated by WISE and {\it GALEX} measurements, respectively. To better show the 60-day lag in the X-ray, we fit the rise caught by optical and X-ray observations with an order 2 polynomial and the plateau with linear fits. 
}
\label{fig:spitz}
\end{figure}

We obtained 17 epochs of follow-up imaging of \tyrion~with the Neil Gehrels {\it Swift} Observatory's \citep{Gehrels2004} Ultraviolet/Optical Telescope (UVOT; \citet{Roming2005,Poole2008}) from 2018 July 30 to 2019 Mar 17 with $2-3$ ks per epoch in the UVW2 filter ($\lambda_{\rm{eff}}$ = 2030 \AA; See Figure~\ref{fig:lc} and~\ref{fig:spitz}).  We detected NUV brightening in the nucleus relative to its archival {\it Galaxy Evolution Explorer} ({\it GALEX}; \citet{Martin2005}) All-Sky Imaging Survey (AIS) magnitude of $NUV = 19.0$ mag (measured with a 6 arcsec radius aperture).  

The source was initially detected with a {\it Swift} UVW2 = 17.7 mag (measured within a 5 arcsec radius aperture), which then faded to UVW2 = 18.0 mag 20 days later, and then remained roughly at that UV flux over the next 50 days.   Note that while some of the UV flux measured by {\it Swift} contains a contribution from extended star-formation (detected in the UV out to a radius of 15 arcsec), the fact that it is variable, and brighter than the archival {\it GALEX} UV central flux indicates that it is associated with the transient. 
The UV-optical color of \tyrion~after subtracting off the {\it GALEX} flux ~is UVW2$-r = -0.45$ mag, very similar to \bco~ (which had NUV$-r = -0.5$ mag, already 0.5 mag bluer than the color range of AGN in both {\it GALEX} and SDSS; \citet{Bianchi2005,Agueros2005}).

\begin{figure*}[ht!] 
\includegraphics[scale=0.5]{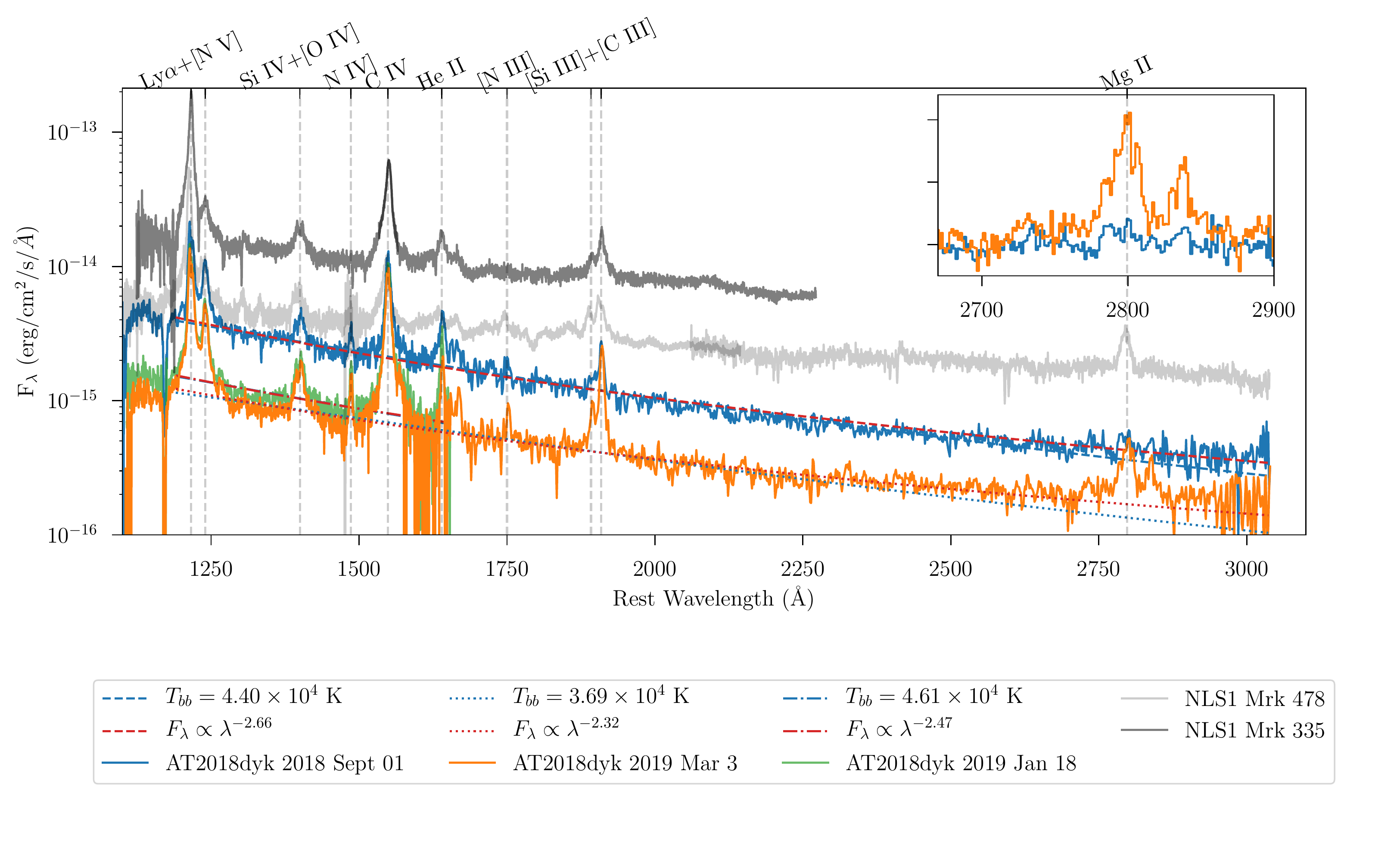}
\caption{{\it HST} UV spectrum of \tyrion~compared to two prototypical NLS1s. Note the presence of high-ionization lines He~II, N~V, O~IV, and C~IV, and the relative weakness of the low-ionization line Mg~II $\lambda$2798 in \tyrion~until later times. {In the second epoch (orange), the continuum of \tyrion~has faded with respect to the first epoch (blue), but broad Mg~II appears to be a lot stronger.}} 
\label{fig:hst}
\end{figure*}

We obtained UV spectroscopy of \tyrion~with the Space Telescope Imaging Spectrograph (STIS) FUV and NUV Multi-Anode Microchannel Array (MAMA) detectors aboard the {\it Hubble Space Telescope} ({\it HST}) for a 2 ks exposure with  0\farcs2 slit width, and G140L ($\lambda$ = 1425 \AA) and G230L ($\lambda$ = 2376 \AA) gratings on 2018 Sept 1, 2019 Jan 18 (only in the FUV\footnote{The second HST epoch had no NUV coverage due to losing lock on the guide stars, and was retaken.}), and 2019 March 3,  shown in Figure~\ref{fig:hst} (Proposal ID: 15331, PI: S.B. Cenko). 

The high spatial resolution of {\it HST} ($\sim 0\farcs5$) enables better isolation of the nuclear emission from the host galaxy light. 
The UV continuum, when masking the emission lines and correcting for Galactic extinction as in Section~\ref{sec:ext}, is an equally good fit to both a blackbody (remaining consistent for both observations within $T=(4.5 \pm 0.3)\times10^4$ K) and a power law with spectral index $\alpha = -2.6\pm0.1$ where $F_\lambda = F_{\lambda,0}\lambda^{\alpha}$ or $\alpha_{\nu} = -\alpha - 2 = 0.6$, with the continuum $F_{\lambda,0}$ decreasing in flux by a factor of 10.7 over 140 days, while the strength of the emission lines remain roughly at the same level.  This blackbody temperature is not unusual for TDEs (e.g. \citealt{vanVelzen2011,Gezari2012,Arcavi2014,Holoien2016a,Holoien2016b,Hung2017}), and the power-law index is within the range of UV slopes observed in quasars \citep[$-1.5 < \alpha_{\nu} < 1.5$;][] {Davis2007}, but steeper than the UV slopes observed in NLS1s \citep[$-2 < \alpha_{\nu} < 0$;][]{Constantin2003}.  
Figure~\ref{fig:hst} shows similarities of the emission features to {\it HST} Faint Object Spectrograph (FOS) spectra of the prototypical NLS1s Mrk 335 and Mrk 478, noting that compared to the NLS1s, the UV spectrum of \tyrion~initially has a weaker low-ionization line Mg II $\lambda$2798, which tends to exhibit weak responsivity in CLAGN (e.g. \citealt{MacLeod2016,Gezari2017}). 
In the latest {\it HST}/STIS epoch, $\sim$6 months after the optical peak, a broad multi-component Mg II line profile appeared, reminiscent of recently ``awakened'' CLAGN Mrk 590 \citep{Mathur2018}. This suggests that a light travel time delay, and not low responsivity, is responsible for Mg II being only marginally detected in the intial observation.  This also implies that Mg II is not co-spatial with the Balmer-line emitting region. 

Galactic extinction has been corrected in these spectra  in the same way as in Section (\ref{sec:ext}), but instead using the UV correction curve for $R_V = 3.1$ given by Eqs. 4.a. and b. in \citet{Cardelli1989}.

\subsection{X-ray}

\begin{figure}[ht!]
\includegraphics[scale=0.33]{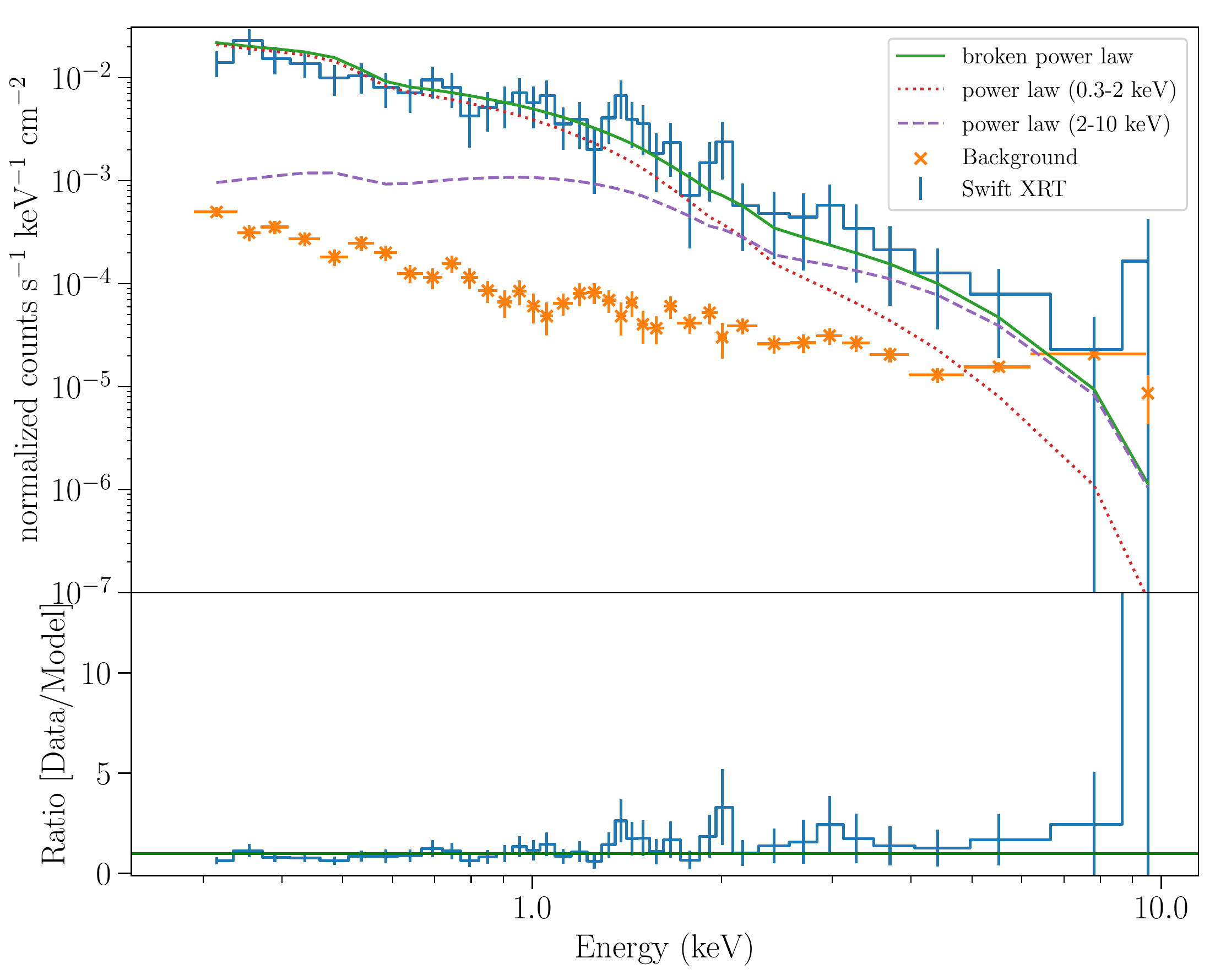}
\includegraphics[scale=0.29]{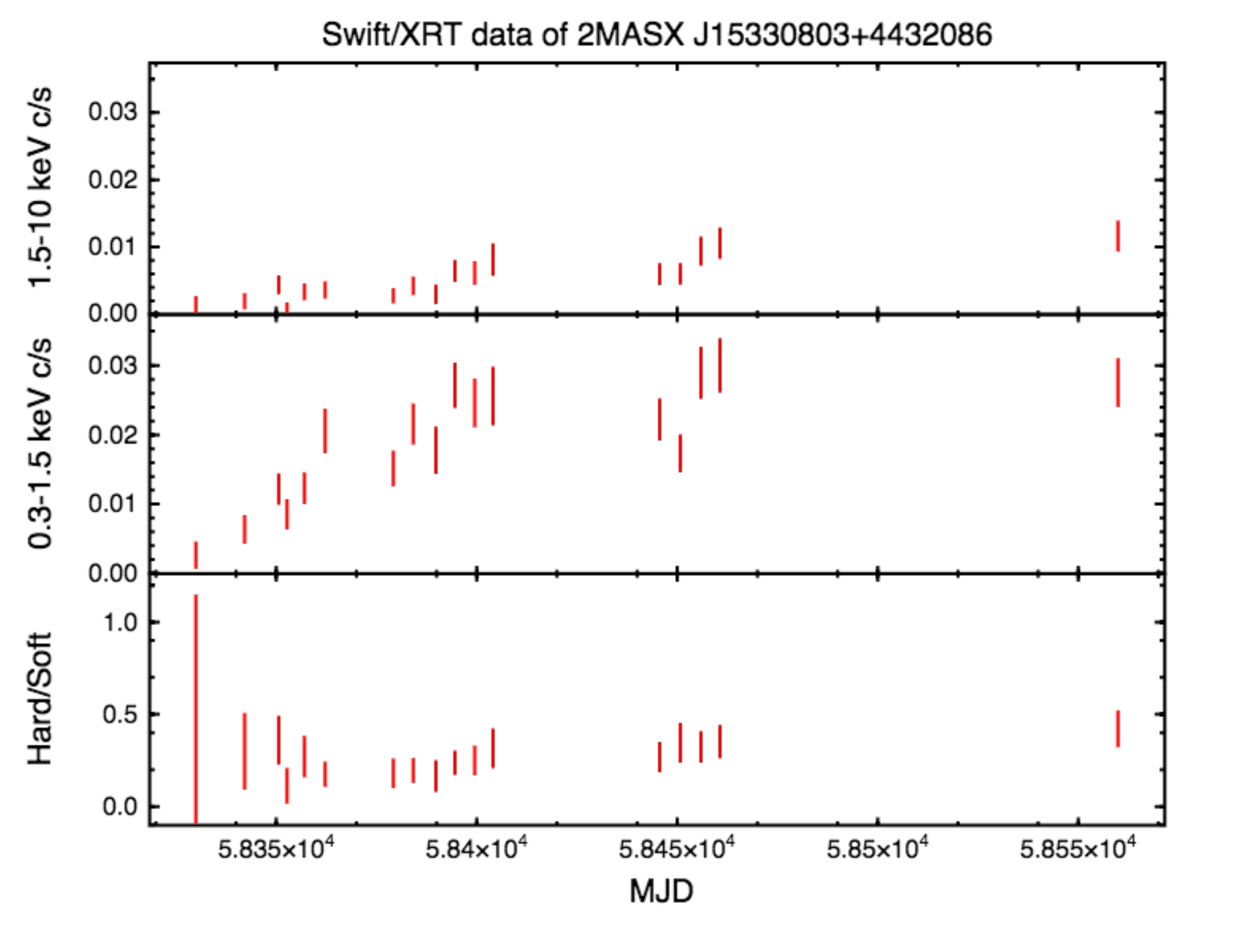}
\caption{Upper panel: XRT spectral fit to a broken power law with soft photon index $\Gamma=2.82^{+0.35}_{-0.26}$ 
described in Section~\ref{sec:x-ray}. Lower panel: Although a slow rise is evident at the 0.01 counts s$^{-1}$ level in the hard band (defined as 1.5$-$10 keV), the hardness ratio light curve shows that the X-ray flare is primarily soft, i.e. 0.3$-$1.5 keV.}
\label{fig:xrt}
\end{figure}

We observed \tyrion~concurrently with 17 exposures of {\it Swift} XRT, detailed in Table~\ref{tab:xray}. 
The XRT data were processed by the XRT Products Page\footnote{\url{http://www.swift.ac.uk/user_objects/}}  \citep{Evans2009} using HEASOFT v6.22\footnote{\url{https://heasarc.gsfc.nasa.gov/docs/software/heasoft/}}. We assessed best-fit models utilizing $\chi^2$ statistics and XSPEC version 12.9.1a \citep{Arnaud1996}. Uncertainties are quoted at 90\% confidence intervals. 
The XRT light curve in the lower panel of Figure~\ref{fig:xrt} shows that \tyrion~is a strongly variable X-ray source, caught rising steadily by an order of magnitude in flux over several months. 
The coadded spectrum (shown in the upper panel of Figure~\ref{fig:xrt}) is well-modeled by a power law with a spectral index of $\Gamma=2.82^{+0.35}_{-0.26}$ and assuming a Galactic extinction of $N_{\rm{H}}=1.76\times10^{20}$ cm$^{-2}$ (computed by the \texttt{NHtot} tool; \citet{Kalberla2005,Schlegel1998}), with no intrinsic absorption and an observed flux between 0.3$-$10 keV of (3.0$\pm0.5)~\times 10^{-13}$ erg cm$^{-2}$ s$^{-1}$. 

\setlength{\tabcolsep}{8pt}
\begin{table*}

\begin{center}
\caption{{\it Swift UVOT/XRT} photometry for \tyrion. Corresponds to  lower panels of Figures~\ref{fig:xrt} and \ref{fig:xmm}. 
}
\label{tab:xray}

\begin{tabular}{cccccccc}
\toprule
      Obs UT & {\it UVOT/XRT} Exposure times &         Count rate &              UVW2 &                Unabsorbed $F_{0.3-10 \rm{keV}}$ &                                                         $L_{\nu,\rm{2~keV}}$ &                                                             $L_{\nu,2500~A}$ &           $\alpha_{\rm{OX}}$ \\
             &                           (s) & ($10^{-2}~s^{-1}$) &          (AB mag) & ($10^{-13}$ erg s$^{-1}$ cm$^{-2}$ \AA$^{-1}$ ) & ($10^{{23}}$ erg s$^{-1}$ {{Hz$^{-1}$}}) & \multicolumn{2}{l}{($10^{{27}}$ erg s$^{-1}$ {{Hz$^{-1}$}})} \\
\midrule
 2018 Jul 30 &                       931/941 &      0.4 $\pm$ 0.3 &  17.72 $\pm$ 0.04 &                                            1.21 &                         {{0.88}} &                         {{8.52}} &  {{-1.91}} \\
 2018 Aug 12 &                      312/2022 &      0.8 $\pm$ 0.3 &  17.90 $\pm$ 0.06 &                                            2.64 &                         {{1.85}} &                         {{7.22}} &  {{-1.76}} \\
 2018 Aug 20 &                      491/3001 &      1.7 $\pm$ 0.3 &  18.05 $\pm$ 0.05 &                                            4.43 &                         {{3.71}} &                         {{6.29}} &  {{-1.62}} \\
 2018 Aug 22 &                      298/2252 &      1.0 $\pm$ 0.2 &  18.03 $\pm$ 0.06 &                                            2.55 &                         {{2.13}} &                         {{6.40}} &  {{-1.72}} \\
 2018 Aug 27 &                      375/3164 &      1.6 $\pm$ 0.3 &  17.99 $\pm$ 0.06 &                                            4.18 &                         {{3.50}} &                         {{6.64}} &  {{-1.64}} \\
 2018 Sep 01 &                      286/2874 &      2.4 $\pm$ 0.3 &  18.05 $\pm$ 0.06 &                                            6.48 &                         {{5.41}} &                         {{6.29}} &  {{-1.56}} \\
 2018 Sep 18 &                      807/3006 &      1.7 $\pm$ 0.3 &  18.10 $\pm$ 0.05 &                                            5.43 &                         {{3.91}} &                         {{6.00}} &  {{-1.61}} \\
 2018 Sep 23 &                      165/3011 &      2.5 $\pm$ 0.3 &  18.05 $\pm$ 0.08 &                                            7.93 &                         {{5.69}} &                         {{6.29}} &  {{-1.55}} \\
 2018 Sep 28 &                      324/1877 &      2.1 $\pm$ 0.4 &  18.20 $\pm$ 0.06 &                                            6.47 &                         {{4.65}} &                         {{5.47}} &  {{-1.56}} \\
 2018 Oct 03 &                      353/3149 &      3.4 $\pm$ 0.4 &  18.08 $\pm$ 0.06 &                                           10.47 &                         {{7.52}} &                         {{6.11}} &  {{-1.50}} \\
 2018 Oct 08 &                      582/2447 &      3.1 $\pm$ 0.4 &  18.18 $\pm$ 0.05 &                                            9.60 &                         {{6.89}} &                         {{5.58}} &  {{-1.50}} \\
 2018 Oct 13 &                     1677/1695 &      3.4 $\pm$ 0.5 &  18.23 $\pm$ 0.04 &                                           10.53 &                         {{7.56}} &                         {{5.33}} &  {{-1.48}} \\
 2018 Nov 23 &                     1329/2931 &      2.8 $\pm$ 0.3 &  18.33 $\pm$ 0.05 &                                            8.68 &                         {{6.22}} &                         {{4.86}} &  {{-1.49}} \\
 2018 Nov 28 &                     1380/2854 &      2.3 $\pm$ 0.3 &  18.36 $\pm$ 0.05 &                                            7.27 &                         {{5.22}} &                         {{4.72}} &  {{-1.52}} \\
 2018 Dec 03 &                     1281/2484 &      3.8 $\pm$ 0.4 &  18.28 $\pm$ 0.05 &                                           11.82 &                         {{8.48}} &                         {{5.09}} &  {{-1.45}} \\
 2018 Dec 08 &                      629/2452 &      4.0 $\pm$ 0.5 &  18.33 $\pm$ 0.05 &                                           12.54 &                         {{9.00}} &                         {{4.86}} &  {{-1.43}} \\
 2019 Mar 17 &                      191/2874 &      3.9 $\pm$ 0.4 &  18.55 $\pm$ 0.09 &                                           12.24 &                         {{8.78}} &                         {{3.97}} &  {{-1.40}} \\
\bottomrule
\end{tabular}
\end{center}
\end{table*}

\begin{figure}[ht!]
\includegraphics[angle=0,scale=.33]{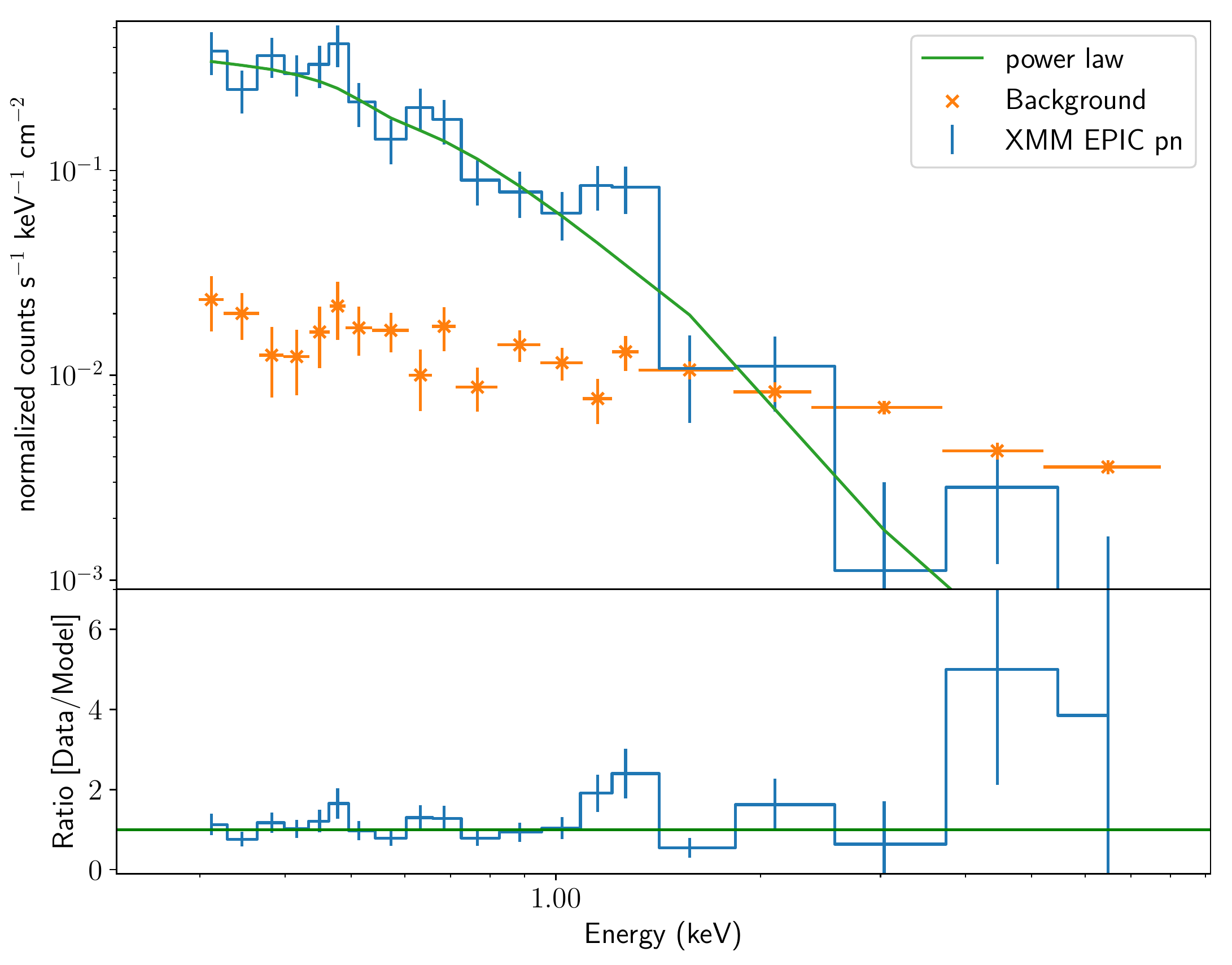}
\includegraphics[scale=0.42]{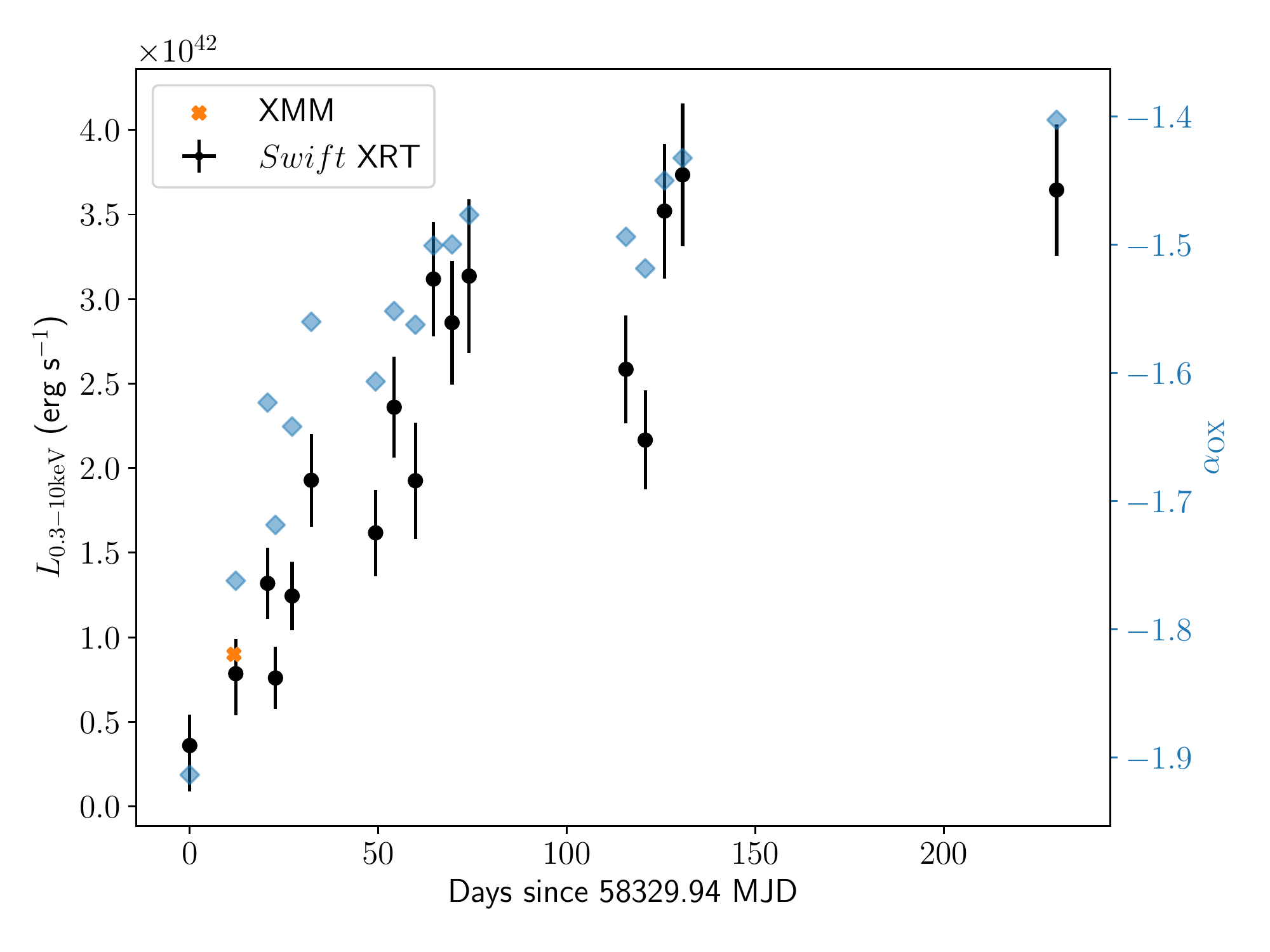}
\caption{Upper panel: The {\it XMM} EPIC pn data of \tyrion~fit to a simple absorbed power law with spectral index $\Gamma=3.02\pm$0.15 shows a prominent, steep soft excess. Lower panel: The X-ray luminosity derived from a power law fit with $\Gamma$=3 is plotted in comparison with $\alpha_{\rm{OX}}$ (described in Section~\ref{sec:x-ray}). 
}
\label{fig:xmm}
\end{figure}

We then observed \tyrion~with the {\it XMM} EPIC pn camera \citep{Struder2001} on 2018 Aug 11 for a 12 ks exposure (Observation ID: 0822040701, PI: S. Gezari). 
We reduced the data using the {\it XMM-Newton} Science Analysis System (SAS) v16.0 \citep{Gabriel2004}. We extracted products with circular source and background (source-free) regions with radii of 35'' and 108'', respectively. To mitigate background flaring and maximize SNR, we filtered {for high background (defined by 13-15 keV) count} rates below 1.75 cts s$^{-1}$. We also adopted CCD event patterns 0 to 4, corresponding to single- and double-pixel events.  We used {\it XMM Newton} EPIC-pn calibration database files updated as of Sept 2018. 
We fit the {\it XMM} EPIC pn data to a simple power law with spectral index $\Gamma = 3.02\pm0.15$ and only Galactic extinction, characteristic of a steep soft excess, and  consistent with the range of photon indices observed for NLS1s \citep[$\overline{\Gamma}=2.8\pm0.9$;][] {Boller1996,Forster1996,Molthagen1998,Rakshit2017}. 

Using the PIMMS count rate calculator\footnote{\url{https://heasarc.gsfc.nasa.gov/cgi-bin/Tools/w3pimms/w3pimms.pl}}, the conversion factor between counts and unabsorbed flux is 3.1$\times 10^{-11}$ for XRT, and 1.5$\times 10^{-12}$ for {\it XMM}.

\subsection{Infrared}
  {\it Spitzer} Infrared Array Camera (IRAC; \citealt{Fazio2004}) observations were triggered for five epochs on 2018 Aug 13 under the approved ToO program (PI: Yan, PID:13251). 
At each epoch, the data were taken for both $3.6$ and $4.5\mu$m, each with a total of 600\,seconds exposure time. A 50 point cycling dither pattern was used. The first three epochal data were taken and used for the analysis when this paper was prepared. The coadded and mosaiced images 
were produced by the standard {\it Spitzer} pipeline and are directly used by our analysis. 

We measured a maximum increase of 0.14 mag compared to archival WISE observations. 
We correct the difference magnitude for the small difference between the bandpass of the two instruments: 0.19, 0.03 mag for channels 1 and 2, respectively, as measured using stars in the field. 
In Figure~\ref{fig:spitz}, we show that this $\nu L_\nu$ at 3.6/4.5 $\mu m$ (subtracting our estimate of the host galaxy baseline as measured by WISE) is greater than $\nu L_\nu$ in the UV, suggesting a large dust covering factor (the fraction of solid angle from the central source obscured by dust). 

NEOWISE data  (WISE, \citealt{Wright2010})  showed there was no variability from the host galaxy of \tyrion~for 1 year prior to its discovery in ZTF, despite the hint of optical variability observed in June 2016 by iPTF (Section~\ref{sec:ztflc}).

\subsection{Radio}


We measure an archival FIRST VLA survey intensity upper limit (including CLEAN bias)  of 0.89 mJy beam$^{-1}$ at the location of the host of \tyrion~in 1997.

\section{Analysis} \label{sec:spc_analysis}

\subsection{Host Galaxy Classification}
\label{sec:liner}
We compare the SDSS spectra of the LINER hosts, observed more than a decade prior to the changing looks caught by ZTF, with follow-up observations taken using the Palomar 60-inch (P60) telescope and the DCT  in Figure~\ref{fig:spc}.  We fit stellar absorption and narrow emission lines to the host spectra with \texttt{pPXF} and results are in Figure~\ref{fig:ppxf}. 
To distinguish them from star-forming galaxies, \citet{Kauffmann2003} define a galaxy as a Seyfert if
\begin{equation*}
    \log([\rm{OIII}]/H\beta) > 0.61/(\log([\rm{NII}]/H\alpha) - 0.05) + 1.3.
\end{equation*}
and \citet{Kewley2001} demarcate a Composite galaxy if 
\begin{equation*}
0.61/(\log([\rm{NII}]/H\alpha) - 0.47) + 1.19 < \log([\rm{OIII}]/H\beta)
\end{equation*}
is true.
 These functions are represented as the dashed and solid lines (respectively) in the BPT 
$[\rm{OIII}]/H\beta$ versus $[\rm{NII}]/H\alpha$ narrow-line diagnostic diagram shown in the upper left panel of Figure~\ref{fig:bpt}. Figure~\ref{fig:bpt} also shows various other line ratio diagnostic diagrams involving the line ratios [OIII]/H$\beta$, [NII]/H$\alpha$, [OI]/H$\alpha$, and [OIII]/[OII] (Baldwin et al. 1981; Kewley et al. 2001, \citet{Kauffmann2003}, Kewley et al. 2006), including the WHAN diagram \citep{CidFernandes2011}, accounting for the equivalent width of H$\alpha$ and the fact that the typical BPT LINER classification contains both ``weak AGN'' and ``retired galaxies'' that have ceased star formation. 

\def\names{{ZTF18aajupnt_sdss_ppxf/\tyrion},{ZTF18aahiqfi_sdss_ppxf/\bronn},{ZTF18aaidlyq_sdss_ppxf/\varys},{\noname_sdss_ppxf/\noname},{ZTF18aasszwr_sdss_ppxf/\podrick},{ZTF18aaabltn_sdss_ppxf/\jorah}}
 \begin{figure*}[!htbp]
\begin{center}
\foreach \name/\subcap in \names {
        \subfigure[\subcap]{
        \includegraphics[width=0.4\textwidth]{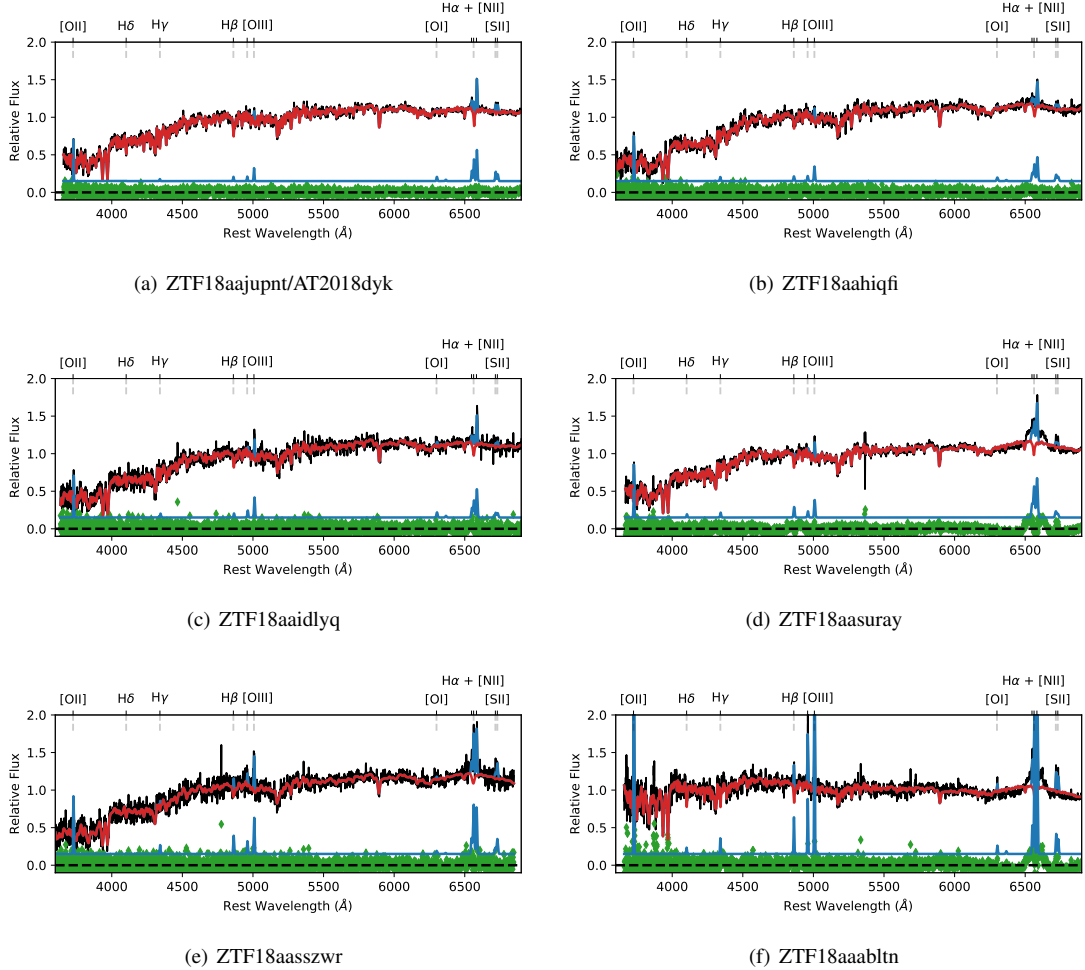}}
}
\end{center}
\caption{The SDSS spectra of the host galaxies were fit using the Penalized Pixel-Fitting (\texttt{PPXF}) method by Cappellari \& Emsellem (2004). Red denotes the stellar population template, blue the emission line fits, and green points the residuals to the total best fit model. Note the poor fit to the [O~II] and [O~III] emission lines of \varys, which are replaced in subsequent analysis by the emission line fits in Figure~18 (available in the electronic version). 
We do not re-analyze \bco~ (not shown here) and instead use the analysis from \citet{Gezari2017}.}\label{fig:ppxf}
\end{figure*}

\begin{figure*}[ht!]
\includegraphics[scale=0.44]{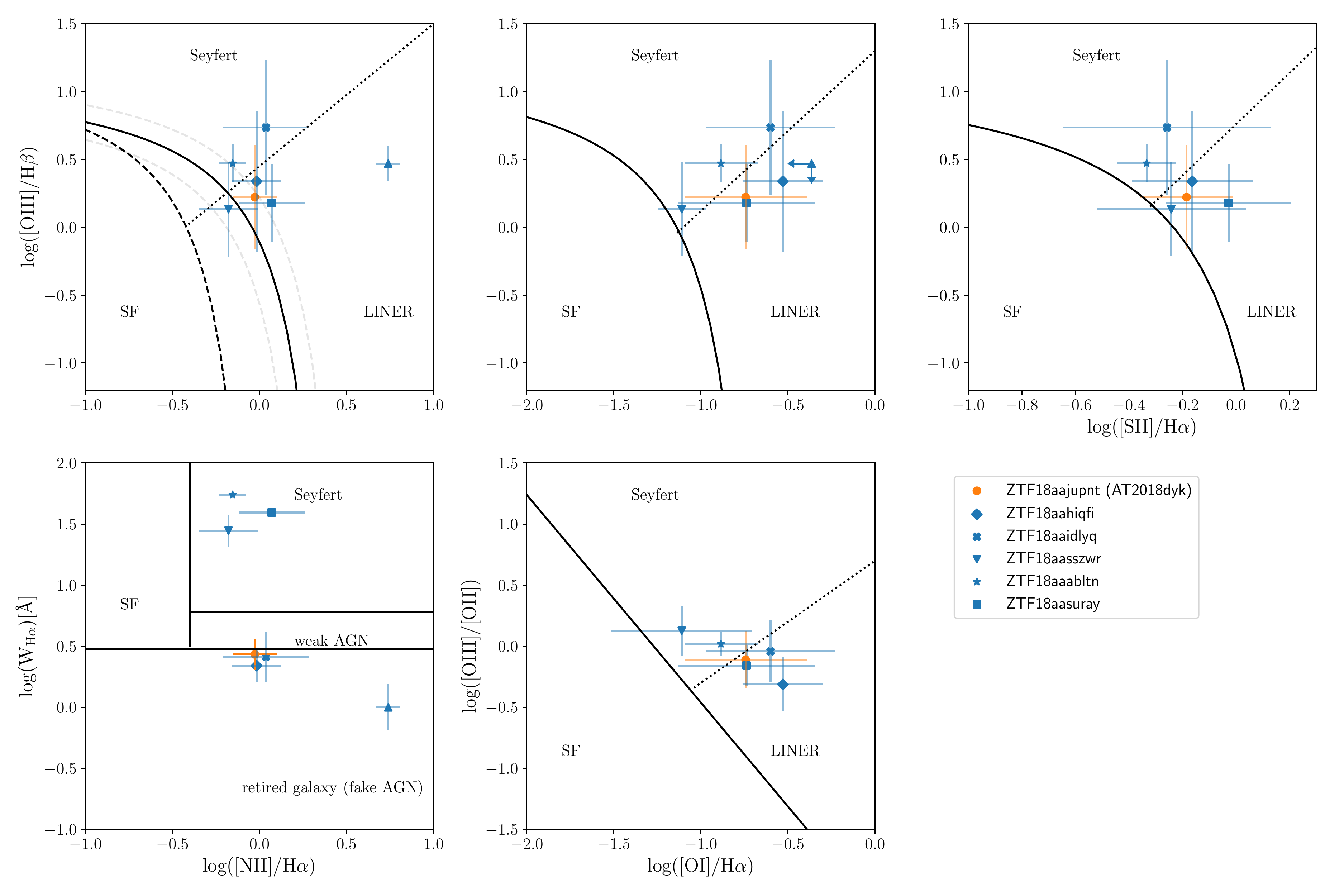}
\caption{Narrow-line diagnostics for the CL LINER sample in the ``off'' state (i.e. their host galaxies), including \bco~ (values from  \citet{Gezari2017}). The majority of the sample is on the borderline between a LINER and Seyfert classification. Note differences in scale.{ }
Upper limits are used when lines are not significantly detected. \newline
 Lower left panel: AGN diagnostic diagram from \citet{CidFernandes2011}. Only three of the sources in the CL LINER sample require a Seyfert to power the Balmer emission lines in their low state, also indicated by the H$\alpha$ line profiles requiring broad components, shown in the fits in Figure~18 (available in the electronic version).}
\label{fig:bpt}
\end{figure*}

Analysis of the archival SDSS spectra of the individual sources in this sample finds that all but CLQ \bco~ exist in the borderline region between LINER and Seyfert classifications for all five diagnostics shown in Figure~\ref{fig:bpt}. According to the diagram of \citet{CidFernandes2011}, both weak and ``fake'' AGN scenarios are plausible within the 1$\sigma$ errorbars for three LINERs in this sample, excluding the host of \bco, which is considered a retired galaxy in this diagnostic, and the hosts of \podrick~and \jorah, which are Seyfert-like (see lower left panel of Figure~\ref{fig:bpt}).

We note that the broad H$\alpha$ component of \jorah~is not completely gone in the spectrum representing its ``off'' state. Although it passed the sample selection criteria of being identified as a LINER in the Portsmouth SDSS DR12 catalog (described in Section~\ref{sec:select}), re-fitting of the line ratios of \jorah~ reveals that it is a Seyfert rather than a LINER. As we measured a broad base in H$\alpha$, we classify it instead as a Sy~1.9 (this is also consistent with prior radio and X-ray detections of this source). \noname~displayed double-peaked broad Balmer emission indicative of a persistent broad line region with unchanging kinematics in both its low and high states. As the peaks did not represent high enough velocities or asymmetric enough profiles to require  separate components, we fit a single broad Gaussian base in this source when measuring the narrow line ratios. Unlike \jorah, those measurements were in agreement with the LINER classification.

Similarly to this work, \citet{Thomas2013} also used \texttt{pPXF} to fit stellar kinematics and the [S~II]/H$\alpha$ ratio diagnostic from \citet{Schawinski2007} (upper right panel of Figure~\ref{fig:bpt}) to classify a source as a LINER; however, they used the Gas and Absorption Line Fitting code (\texttt{GANDALF} v1.5; \citet{Sarzi2017}) to fit emission lines, whereas we use a simple multi-component Gaussian profile fit to the narrow lines in the stellar-template-subtracted spectra (for these model fits, see Figure~18 in the Appendix available in the electronic version). 
There may also be a discrepancy stemming from \texttt{GANDALF} correcting for dust\textemdash the majority of this sample have Balmer decrement $f_{\rm{H\alpha}}/f_{\rm{H\beta}} > 3.1$, indicative of strong intrinsic reddening. However, we choose not to apply a dust correction since it is an uncertain measurement for this sample, due to the weak emission line intensities. 
 
\subsection{Black Hole Masses} 

\label{sec:mbh}
In order to shed light on the physical differences between the individual AGN in this sample, we estimate the black hole masses of the CLAGN hosts using several methods. The broad H$\beta$ line is the most common virial estimator for BH masses at low redshift ($z \lesssim 0.4$; e.g. \citealt{Marziani2012}).

\begin{equation*}
M_{\rm{BH},vir}=1.5\times10^5 \left (\frac{R_{\rm{{BLR}}}}{\rm{light~days}} \right ) \left (\frac{\rm{FWHM(H\beta)}}{10^3\rm{km s^{-1}}} \right)^2 M_{\odot}
\end{equation*}

\noindent where $R_{\rm{{BLR}}}=32.9(\frac{\lambda L_{5100A}}{10^{44}\rm{erg s^{-1}}})^{0.7}$ light days \citep{Kaspi2000}. 
We also calculate $M_{\rm{BH}}$ from the host galaxy luminosity following \citet{McLure2002} such that

\begin{equation*}
M_{\rm{BH},M_r}=-0.5M_{r,\rm{host}} - 2.96,
\end{equation*}

\noindent the host bulge stellar mass using the relation from \citet{Haring2004}

\begin{equation*}
{\rm log}(M_{\rm{BH},Bulge}[M_\odot])=\rm{log}(0.0014 M_{\rm{Bulge}}[M_\odot]), 
\end{equation*}

\noindent and from the stellar velocity dispersion ($\sigma_\bigstar${;} measured from the SDSS spectrum using the \texttt{pPXF} method) using the $M_{\rm{BH}}-\sigma$ relation from \citet{Tremaine2002}

\begin{equation*}
{\rm log} M_{\rm{BH},\sigma\bigstar}[M_\odot] = 8.13 + 4.02  \rm{log}(\sigma_{\bigstar}/200~{\rm km~s}^{-1}).
\end{equation*}

The results of these measurements are summarized in Table~\ref{tab:mbh}, and discussed further in Section~\ref{sec:tdes}.

\subsection{Comparison to Tidal Disruption Events}
\label{sec:tdes}

It is important to compare the properties of this class of AGN ``turning-on'' from quiescence with a related phenomenon of tidal disruption events (TDEs).  When a star passes close enough to a central black hole to be ripped apart by tidal forces, roughly half of the stellar debris will remain bound to the black hole and provide a fresh supply of gas to accrete onto the black hole.  The evolution of the flare of radiation from a TDE is regulated by the fallback timescale ($t_{\rm fb}$), the time delay for the most tightly bound debris to return to pericenter after disruption, and the circularization timescale, which is dependent on the efficiency at which the debris streams shock and circularize due to general relativistic precession.  Interestingly, the virial black hole mass for all the CL LINERS in the iPTF/ZTF sample are above the black hole mass for which a solar-type star can be disrupted outside the event horizon ($M_{\rm{BH}}\lesssim10^{8}~M_{\odot}$).  The only exception is \tyrion, which as a NLS1 in its ``on'' state, thus with narrower lines, naturally implies a smaller black hole mass {for equal luminosity} with this method.  However, the black hole mass inferred from the host galaxy velocity dispersion and bulge mass suggest a larger black hole mass of log$(M_{\rm BH}/M_\odot) = 7.6-7.8$.  
This trend of the black hole mass from the virial method being much smaller is consistent with the work of \citet{Rakshit2017}, who suggest that the smaller Balmer line widths measured in NLS1s which lead to lower BH masses are due to the geometrical effects of being viewed more face-on ($\langle i\rangle=26\degree$) compared to normal broad line Sy~1s ($\langle i\rangle=41\degree$). This claim is backed up by spectropolarimetric studies of NLS1s \citep{Baldi2016}, 
Alternately, \citet{Marconi2008} suggested that in rapidly accreting objects (including NLS1s), enhanced ionizing radiation pressure could also lead to underestimates of virial black hole mass estimates.  

It is also possible that these transitioning AGN do not obey the radius-luminosity relation established from reverberation mapping studies of Seyfert galaxies.  If we instead use the black hole mass estimates from the host galaxy velocity dispersion, luminosity, and/or stellar mass, we find that the CL LINER sample have black hole masses of log$(M_{\rm BH}/M_\odot) \sim 7-8$, close to, but not necessarily exceeding the upper mass limit for the tidal disruption of a solar-type star.  

\begin{figure*}[!htbp]
\centering
\includegraphics[scale=0.7]{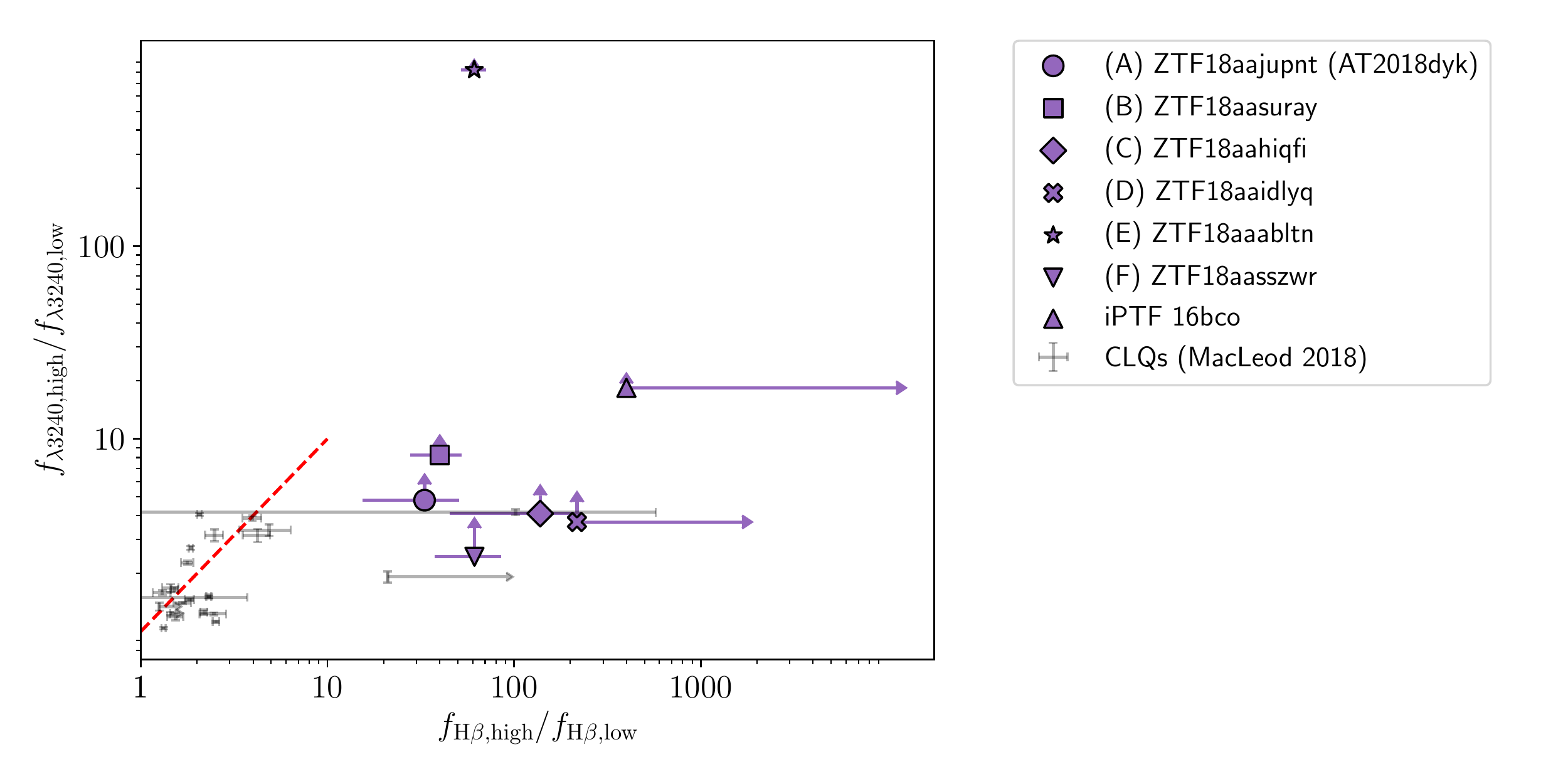}
\caption{Ratio of continuum flux change as a function of broad line flux change for our changing-look LINER sample (filled shapes) in comparison to changing-look Seyferts.  \tyrion~(purple circle), is intermediate in flux ratio and H$\beta$ ratio space between Seyfert CLAGN (black, lower left) and the other CL LINERs in this sample. The red dotted line denotes a 1:1 ratio between the continuum and H$\beta$ fluxes. \bco, \noname, \podrick, \bco~are outliers in differential continuum space (although we collected spectra of the latter two with an IFU spectrograph that can be unreliable at bluer wavelengths), and \bco, \varys, and \bronn~are outliers in H$\beta$ luminosity space compared to that of the Seyfert CLAGN. All have much larger (by a factor of $>10$) changes in broad line flux than the changing-look Seyfert sample. 
The $f_{\lambda3240}$ ratio measurements are represented as lower limits, as there is stellar contamination in the low (LINER) state. For sources with H$\beta$ undetected in the low state, the errorbars on the lower limits are at the 1-$\sigma$ level. Adapted from Figure 6 in  \citet{MacLeod2019}. 
}
\label{fig:macleod}
\end{figure*}

We can also compare the light curves and spectra of our CL LINERs to TDEs.  The quiescence in the light curves before the onset of their flaring activity, their blue colors ($g-r < 0$) during the flaring in most of the cases, as well as their smooth decline from peak are generally consistent with the TDE scenario.  The main distinction is in their spectral properties at peak.  The five objects caught transitioning from a LINER to a type 1 AGN show spectra in their ``on'' state that are almost indistinguishable from normal quasars, besides the relative weakness of [O~III].  In contrast, TDEs exhibit exclusively broad emission lines;  broad He~II $\lambda4686$ emission, and/or broad H$\alpha$ and H$\beta$ lines, and sometimes broad He I, but with line luminosities of $\lesssim 10^{41}$ ergs s$^{-1}$ \citep{Arcavi2014,Brown2016, Hung2017, Holoien2018}, well below the CL LINERs (see Figure~\ref{fig:macleod}).  Furthermore, the X-ray spectra of the CL LINERs with X-ray observations in their ``on'' state, \bco~ and \tyrion, are well described by a power-law, with $\Gamma = 2.1$ \citep{Gezari2017} and $\Gamma = 3.0$, respectively, and are clearly distinct from the extremely soft blackbody spectra with $kT \sim 0.04-0.10$ keV characteristic of both optically and X-ray selected TDEs \citep{Komossa2002, Miller2015,VanVelzen2019}. 
We present a more detailed comparison of the observed properties of \tyrion~with TDEs in Section \ref{sec:tyrion}.

\subsection{Comparison to Seyfert CLAGN} \label{sec:macleod}
We measure the H$\alpha$ and [O~III] $\lambda$5007 luminosities for this sample in their ``on'' state in Figure~\ref{fig:haoiii} and compare to that of SDSS Sy~1s (including NLS1s; \citet{Mullaney2013}) and quasars \citep{Shen2011}. All AGN in this sample display [O~III] $\lambda$5007 luminosities significantly below average for their observed broad H$\alpha$ luminosity in their ``on'' state, consistent with the findings of \citet{Gezari2017}, that CLQs with appearing (disappearing) broadlines were in general closer to the fringe (average) of the quasar distribution. However, for  \podrick~and \jorah, only upper limits of [O~III] were possible due to the low SNR for narrow lines of the low-resolution ($R\sim100$) follow up spectra.

\citet{MacLeod2019} systematically obtained spectra for highly-variable candidate CLQs (defined as type 1 AGN transitioning to type 2s or vice versa) within the SDSS footprint, requiring Pan-STARRS 1 variability exceeding $|\Delta g|>1$ mag and $|\Delta r|>0.5$ mag. We find agreement with their measured positive correlation between broad emission line and continuum flux changes, but find that our sample of CL LINERs is more extreme in the parameter space of continuum and H$\beta$ flux ratios (ranging from 2$-$800 and 12$-$400, respectively) than the CLQ sample from \citet{MacLeod2019} (with $f_{\rm{high}}/f_{\rm{low}} = 1-7$ and 2$-$8 for continuum and H$\beta$, respectively), shown in Figure~\ref{fig:macleod}. {Although the range of redshifts of the two samples differ, we confirm through a comparison with measurements of published local Seyfert CLAGN that their continuum and H$\beta$ ratios are consistent with that of the CLQ sample.} When rest-frame flux at 3240 \AA~was not available to us due to inconsistent spectral coverage, we measured flux at the shortest available comparable wavelength. 

\subsection{Eddington Ratio Estimates}
\label{sec:ledd}

We compute the Eddington ratio ($L_{\rm{bol}}/L_{\rm{Edd}}$) 
for the sample in their ``on'' state assuming $L_{\rm{bol}} = 9\lambda L_{5100A}$ \citep{Kaspi2000}, summarized in the final column of Table~\ref{tab:mbh}. 
$L_{\rm{bol}}$ in the ``on'' state is measured using difference imaging in the filter with central wavelength closest to rest-frame 5100~\AA~for each source ($r$-band for higher-redshift sources \bco, \varys, and \podrick, and $g$-band for all others). 
$L_{\rm{bol}}$ in the ``off'' state is measured from the reddening corrected  $L_{\rm{[O~III]}}$ narrow line luminosity correlation to $L_{\rm{2-10~keV}}$ for type 2 AGN (Equation 1 in \citet{Netzer2006}) and using the bolometric correction {for LINER-like AGN from \citet{Ho2009}}, $L_{\rm{bol}}=15.8 L_X$. 
{We confirm that the reported luminosities are robust to systematics introduced by our choice of the bolometric corrections by computing $L_{\rm{bol}}$ in the high state for those sources with available $L_{\rm{2-10~keV}}$ measurements, and find that the two methods are consistent within a factor of $\sim4$.}

While virial black hole masses based on the broad H$\beta$ line and continuum luminosity are more generally used for AGN, those relations are based on reverberation mapping studies which were never done specifically for NLS1s. 
Thus, for the remainder of this work, we adopt BH mass estimates for the sample to be consistent with $M_{\rm{BH}}$ from stellar velocity dispersions as described in Section~\ref{sec:mbh} and summarized in Table~\ref{tab:mbh}. 

The Seyfert CLAGN with appearing broad emission lines in the variability-selected \citet{MacLeod2019} sample (summarized in Section~\ref{sec:macleod}) have $-2 \lesssim$ log($L/L_{\rm{Edd}}$) $\lesssim -1$, slightly below that of a control sample of extremely variable quasars and normal SDSS DR7 quasars. For the range of this small sample {($-2.7 \lesssim$ log($L/L_{\rm{Edd}}$) $\lesssim -1.2$), the Eddington ratios of the CL LINERs are well matched to the population of CLQs} in their ``on'' state. The corresponding upper limits of log $(L/L_{\rm{Edd}}) < -2$ in the ``off'' states of the LINER host galaxies are in good agreement with that of the \citet{MacLeod2019} CL population that has dimmed.



\begin{figure*}
\includegraphics[scale=0.65]{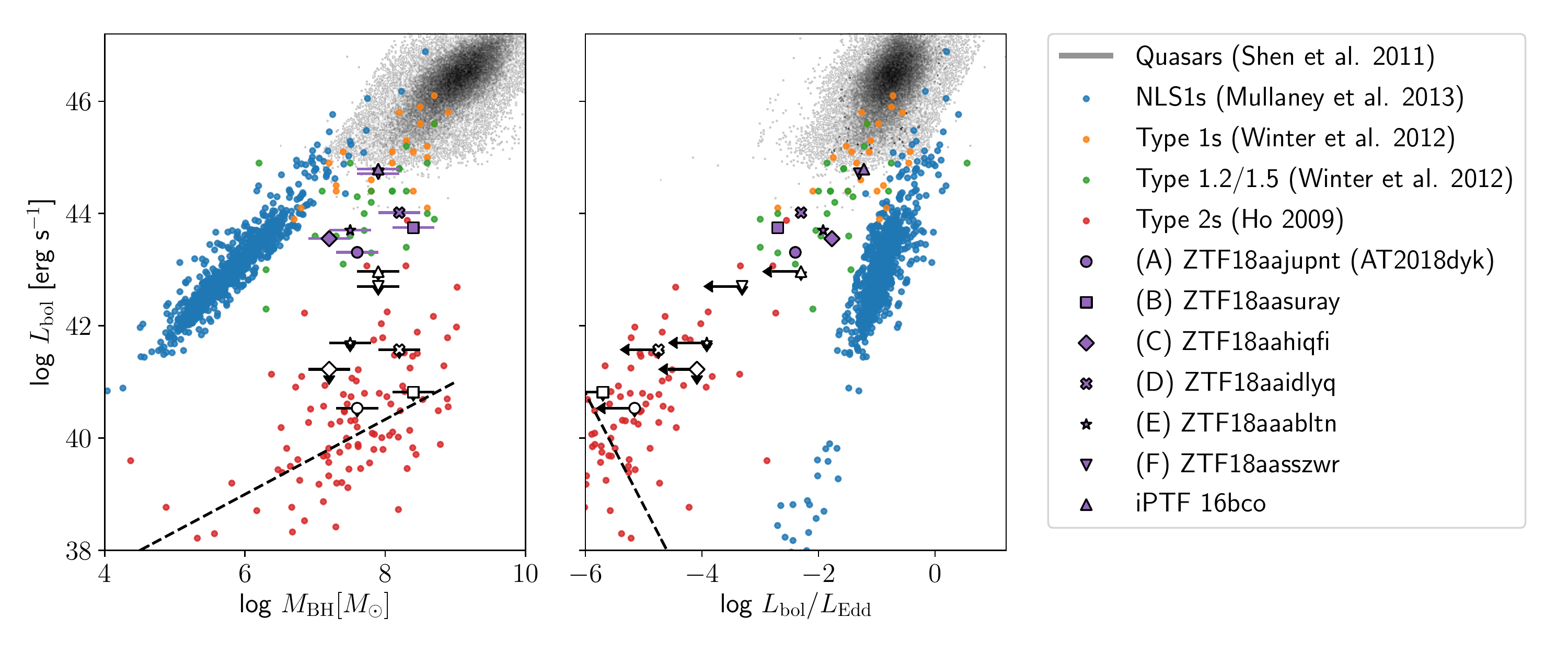}
\caption{AGN, when separated by spectral classification, show the rough evolutionary sequence in parameter space of black hole mass $M_{\rm{BH}}$, bolometric luminosity $L_{\rm{bol}}$, and Eddington ratio $L_{\rm{bol}}/L_{\rm{Edd}}$ described in Section~\ref{sec:ledd}. The dotted lines denote the critical values above which a BLR can be sustained from \citet{Elitzur2009} described in Section~\ref{sec:ledd}, to which we compare the measured values for the sample (filled, purple shapes with same mapping as in Figure~\ref{fig:macleod}) and their hosts (unfilled, black{, primarily upper limits}). The CL LINER sample in their ``on'' state is consistent with the type 1s (orange points){ and intermediate types (green points)}. We note that the type 2 sample from \citet{Ho2009} contains LINER2s and LLAGN. Adapted from Figure 1 in \citet{Elitzur2014}. {The errorbars on bolometric luminosity are comparable to the size of the points.} 
}
\label{fig:elitzur}
\end{figure*}

\citet{Elitzur2014} predict a natural sequence within the disk-wind scenario in which AGN evolve from displaying to lacking broad optical emission lines. This evolution is driven by variations in accretion rate (with the critical value parameterized by $L_{\rm bol}/M^{2/3}_{\rm{BH}}$), as well as the availability of ionizing radiation from the central engine. The BLR is therefore posited to be assembled following an increase in accretion rate (likely due to  instabilities to match the fast timescales observed; \citet{Rumbaugh2018}). 
Due to an insufficient cloud flow rate and lack of ionizing photons, no BLR can be sustained below the critical accretion rate or bolometric luminosity ($L_{\rm{bol}} \leq 5 \times 10^{39} M^{2/3}_7$ erg s$^{-1}$, \citet{Elitzur2009}). This spectral evolutionary pathway is supported by modeling an SDSS-selected sample of Seyferts of various types and spanning $L/L_{\rm{Edd}}\sim10^{-3}$ to 0  \citep{SternLaor2012}, for which accretion rate progressively decreased with luminosity from type 1s to type 2-like AGN. 
In Figure~\ref{fig:elitzur} we recreate this sequence represented by AGN with different spectral classifications occupying distinct regions of the $L_{\rm{bol}}-M_{\rm{BH}}-L/L_{\rm{Edd}}$ parameter space and roughly separated by the critical threshold of \citet{Elitzur2009}. We overplot the CL LINER sample in their ``on'' states which overlap roughly with the {Seyfert type 1 and intermediate type} sources, and in the ``off'' LINER states which overlap largely with the type 2s and border on the intermediate type 1.2/1.5 Seyferts. 

The bolometric luminosities (and therefore the Eddington ratios) are upper limits in Figure~\ref{fig:elitzur} due to the ``off'' spectra being almost entirely host dominated. 
\bco, \noname, \varys, and \podrick~approach the quasars in their ``on'' states, and \tyrion~does not fall squarely among the NLS1s but instead in the border region between types.  
The least luminous sources in the sample, \tyrion~and \noname,  approach most closely the critical Eddington ratio for the existence of a BLR in their ``off'' states, and the most luminous \bco~ is closest to the intermediate types in its LINER state.

\subsection{\tyrion: A LINER Changing-Look to a NLS1}
\label{sec:tyrion}
For the following analysis we focus on \tyrion, for which we have the most extensive follow-up data, and which showed the appearance of coronal lines along with X-ray variability. 
The difference imaging light curve of this event displays a plateau similar to that of \bco~ (\citet{Gezari2017}; see comparison in Figure~\ref{fig:16bco}), before fading gradually over several months in a manner similar to that of CL LINER \podrick, rather than the power-law decline characteristic of an optical TDE light curve (e.g. \citealt{Hung2018}). 

The lack of IR variability in NEOWISE leading up to the turn-on of \tyrion~constrains the presence of any IR AGN activity or dust echo in this host to $< 10$ months. 
W1-W2 is never greater than $\sim$0.02 during this time, far below the 0.8 threshold AGN diagnostic value from \citet{Stern2012}. 
Stability in the CRTS light curve similarly confirms that no AGN-like variability was present for 13 years prior to its discovery with ZTF. 
There was, however, a hint of some $\sim$0.1 mag flaring in the CRTS light curve in June 2006 and April 2007. 
Additionally, we extracted forced photometry \citep{Masci2017} for \tyrion~from the PTF database covering June 2011 to June 2016, and there were only 8 marginal detections near the limiting magnitude of PTF (from 20 to 20.9 $r$-band mag) for the last 15 days of this range.

To reproduce the photometry of \tyrion, any physical explanation must explain a rise time of $\sim$50 days and a slow decline rate of $\sim$0.5 mag in 60 days, both quite unusual for a TDE or supernova (e.g. \citealt{VanVelzen2019}).  \citet{Arcavi2018} note that the difference imaging light curve of \tyrion~peaks at an absolute magnitude of $-17.4$ mag, which is much fainter than the majority of TDEs by several magnitudes, excluding iPTF16fnl \citep{Blagorodnova2017}. A power law and blackbody give nearly identical fits to the UV spectra (with $\overline{T_{\rm{bb}}}=4.5 \times 10^4$ K) with no change in the slope as the continuum fades over $\sim$140 days; Figure~\ref{fig:hst}). The optical continuum in Figure~\ref{fig:resid} is well fitted with a power-law, consistent with the Rayleigh-Jeans tail of a blackbody.

In the UV, the observed spectrum does not resemble that of a TDE in a LINER (e.g. ASASSN-14li, \citet{Cenko2016}).  
Instead, the UV spectrum of \tyrion~is very similar to the UV spectra of normal NLS1s, with a similar spectral slope and peaked, broad emission line shapes (see Figure~\ref{fig:hst}).  In particular, \tyrion~has a strong C~IV $\lambda\lambda 1548, 1551$ line and C~III] $\lambda 1909$ line, which is typical of NLS1s, but not detected in all the TDEs with HST UV spectra: ASASSN-14li \citep{Cenko2016}, iPTF15af \citep{Blagorodnova2019}, iPTF16fnl \citep{Brown2018}, AT2018zr \citep{Hung2019}.  Interestingly though, \tyrion~does show N IV] $\lambda 1486$ emission, which is just barely detected in NLS1s \citep{Constantin2003} and {\it is} detected in the UV spectrum of TDE ASASSN-14li, which was argued to be N-rich. The critical density $3.4\times 10^{10}$ cm$^{-3}$ of the intercombination N IV] $\lambda 1486$ line provides an upper limit to the density of this gas in \tyrion~\citep{Nussbaumer1979}. The late-time increase in the Mg II line has not been detected in a TDE; in fact the opposite trend has potentially been observed: the brightening of broad Mg II with the fading of the transient in TDE AT2018zr \citep{Hung2019}.  Finally, \tyrion~demonstrates none of the broad absorption features seen in the UV spectra of TDEs, and has been associated with powerful outflows launched by the accretion process in a TDE.

\subsubsection{Coronal Line Emission from \tyrion}
\label{sec:tyrion_cl}
We report line measurements of the Keck spectrum of \tyrion~in Table~\ref{tab:lines}. We choose to analyze the spectrum from this instrument because of its sufficiently high SNR and spectral resolution to measure the presence of coronal lines. For each of these measurements, the stellar population of the host galaxy represented by the {\texttt ppxf} fit has been subtracted (See Figure~\ref{fig:ppxf} for a visual of the stellar model template). 

\begin{figure*}[ht!]
\includegraphics[scale=0.6]{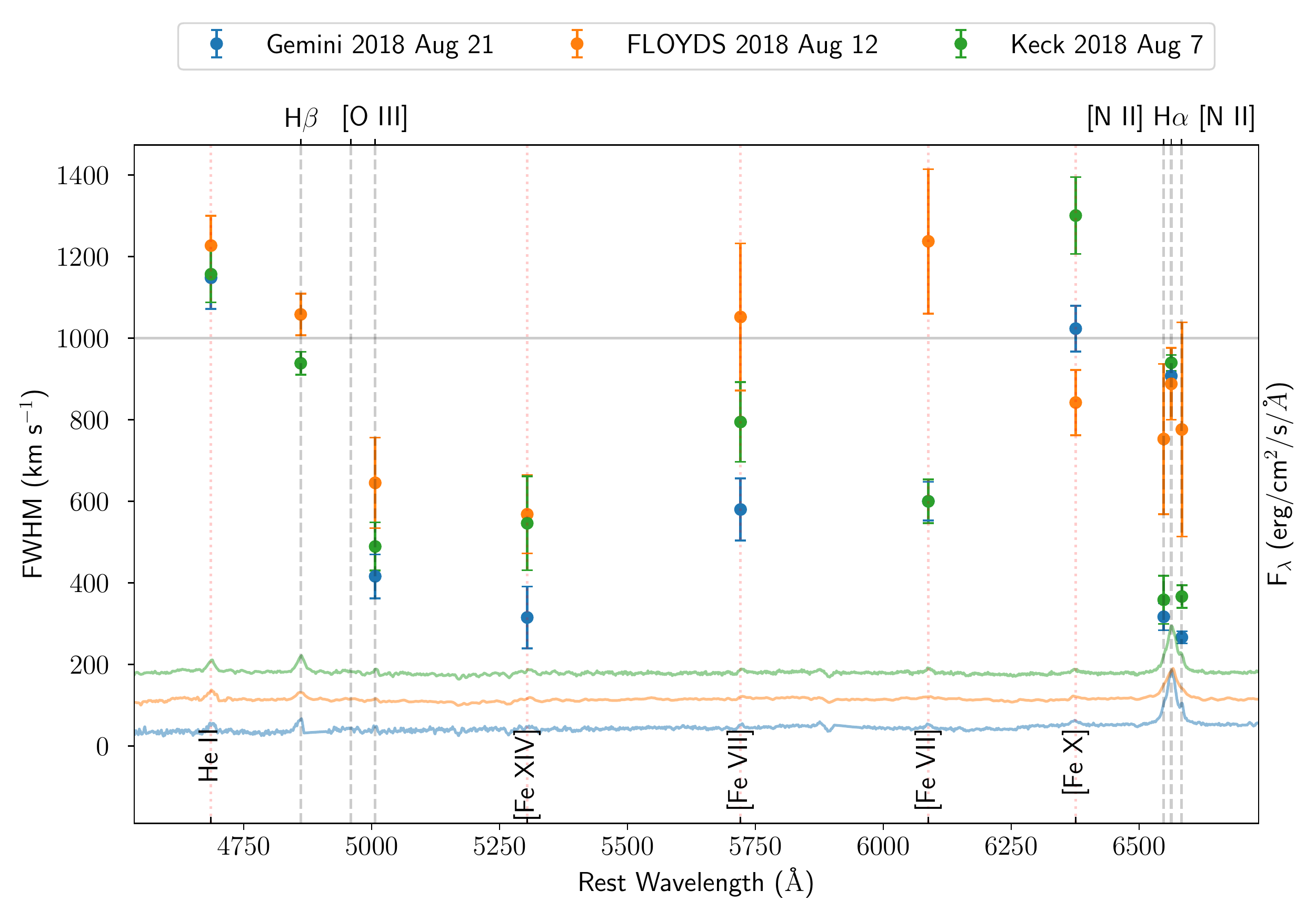}
\caption{FWHM of H$\alpha$, H$\beta$, and the coronal lines for each high-resolution optical observation of \tyrion~in its ``on" state. The stellar population of the host galaxy has been subtracted.}
\label{fig:fwhm}
\end{figure*}

 The width of the majority of the coronal lines is narrower than the widths of the broad permitted AGN emission lines (see Figure~\ref{fig:fwhm}), as is expected from forbidden high-ionization collisionally-excited emission because it originates from a larger distance from the ionization source. However, there is no strong evidence that the coronal emission lines in \tyrion~are observed with widths between the BL and NL emission, as expected in the scenario in which gas is outflowing from an intermediate coronal line region (CLR; e.g. \citealt{Mullaney2008}). The [Fe~X] line is unlikely to be broader than expected due to blending with the [O~I] $\lambda$6364 line (e.g. \citealt{Pelat1987}), as it is in a 1:3 ratio with the [O~I] $\lambda$6300 line which is observed to be weaker than [Fe~X] in this source. In Sy~1s, [Fe~X] tends to be relatively stronger than the other coronal lines (e.g. \citealt{Pfeiffer2000}). However, in Seyferts the CL emission is typically measured to be only a few percent of the strength of [O~III] $\lambda$5007).
 
 The fact that [Fe~X] $\lambda$6374 is stronger than [O~III] $\lambda$5007 places \tyrion~away from other Seyferts and instead among the $<$10 known extreme coronal line emitters (ECLEs) in this parameter space. We discuss further the ECLE scenario in Section~\ref{sec:t_vs_a}. We note that the weakness of [O~III] may be due to light travel time effects, and thus may strengthen with time. 
 
We note the significant spectral differences between \tyrion~and SN 2005ip post-peak \citep{Smith2009}. SN 2005ip has much more prominent coronal lines than even the example ECLEs, as well as a strong hydrogen emission series, much broader than the quasar \bco~ plotted alongside it.
 
 \citet{Korista1998} presented a model by which coronal lines are the result of ISM interaction with bare Seyfert nuclei, i.e. AGN lacking any X-ray/UV evidence of intrinsic absorption by ionized gas along the line of sight to the AGN. This model is consistent with our finding of no intrinsic absorption in the X-ray spectra of \tyrion. 

\subsubsection{\tyrion~as a NLS1 in its ``On'' State}
\label{sec:nls1}

At the other extreme of eigenvectors of AGN spectral properties are narrow-line Seyfert 1 galaxies (NLS1s), a subclass of AGN that are characterized by relatively narrow Balmer lines (FHWM $<$ 2000 km s$^{-1}$), strong broad Fe~II emission, [O~III] $\lambda$5007/H$\beta_{\rm{tot}} < 3$, a prominent soft X-ray excess (e.g. \citealt{Puchnarewicz1992}), and dramatic variability, especially in the X-rays \cite[e.g.][]{Pogge2000,Frederick2018}. 
These spectral properties of NLS1s are attributed to
lower-mass central black holes (5 $<$ log($M_{\rm{BH}}[M_\odot]$) $<$ 8; e.g. \citealt{Mathur2001}) that are thought to accrete at high Eddington ratios \citep{Pounds1995,Wang1996,Grupe2010,Xu2012}. 

We measure 1000 $\lesssim$ FWHM(H$\beta$) $<$ 2000 km s$^{-1}$  which is indicative of a narrow-line Seyfert 1 galaxy in the AGN interpretation \citep{Goodrich1989}, as well as the fact that the Balmer lines are significantly better fits to Lorentzian line profiles than Gaussians \citep{Nikolajuk2009}. However, the FWHM limits between Sy~2s, NLS1s and Sy~1s is somewhat arbitrary \citep{VeronCetty2001,Mullaney2008}, and may even be better set at 2200 km s$^{-1}$ \citep{Rakshit2017}. The fact that some of the line measurements fall short of this cutoff could speak to the intermediate nature of this transitioning object in the changing-look scenario. The virial mass measurement for \tyrion~is consistent with the NLS1 interpretation, as NLS1s display properties consistent with AGNs with lower masses \citep{Grupe2004}, though it is toward the high end of the NLS1 mass distribution \citep{Xu2012}. 
Also consistent with the NLS1 scenario is that [O  III] $\lambda$5007 / H$\beta = 0.1 < 3$ \citep{Osterbrock1985}. 
However, [O  III] $\lambda$5007 appears to be relatively quite weak when compared to that of of prototypical NLS1, Mrk 618, in Figure~\ref{fig:tyrion_spc}. 
It should also be noted that the coronal lines in \tyrion~appear to be symmetric and at the same systematic redshift as the Balmer series and low-ionization forbidden lines, whereas coronal lines in Seyferts can be significantly broadened, asymmetric, and blueshifted consistent with an outflowing wind launched between the BLR and NLR (\citet{RodriguezArdila2006}; in NLS1s: \citet{Erkens1997,Mullaney2008,Porquet1999}). 
This is less common, but not unheard of, for ECLEs (See Section~\ref{sec:t_vs_a}).

\begin{figure*}[ht!]
\plotone{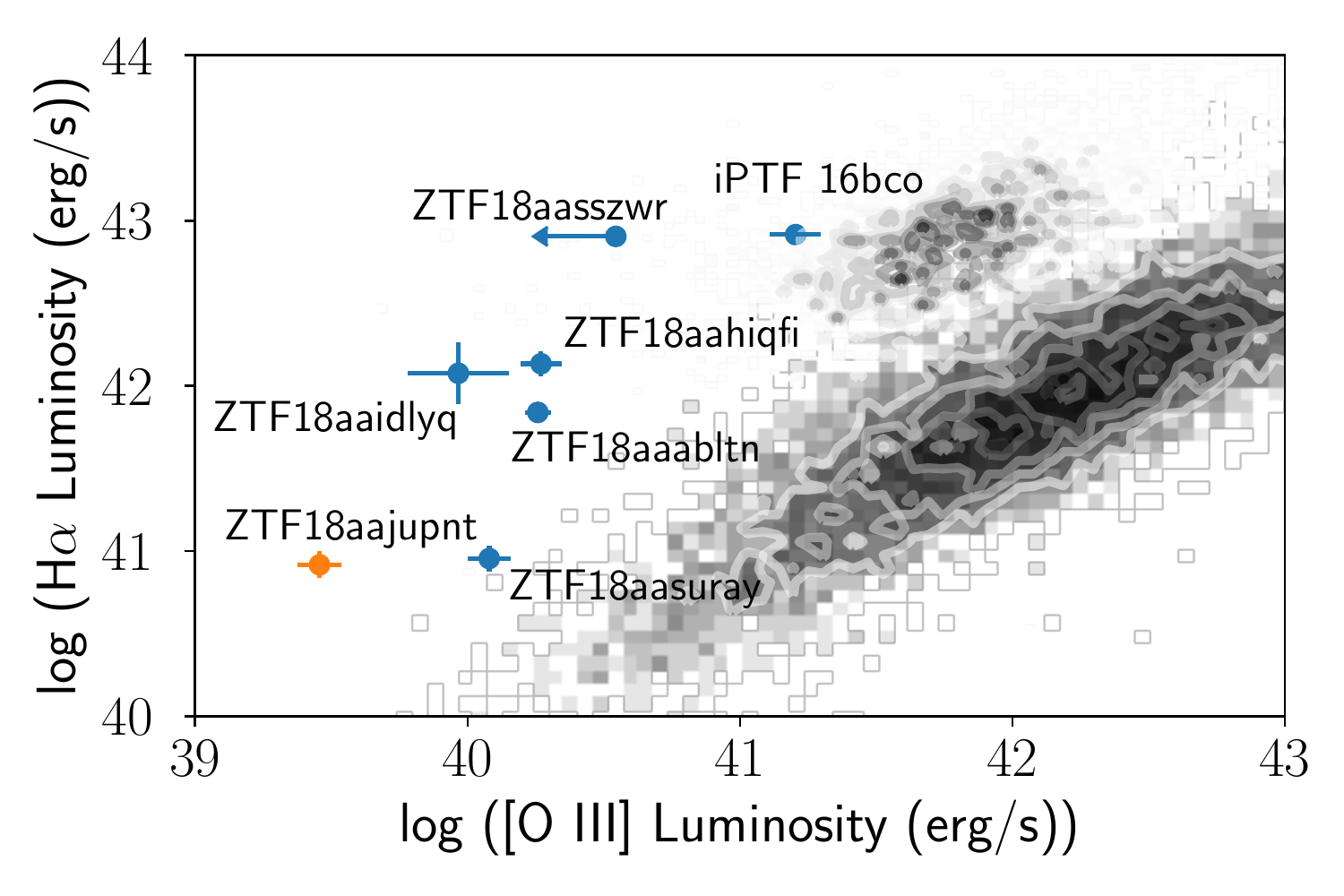}
\caption{H$\alpha$ and [O~III] $\lambda$5007 line luminosities measured for this sample of CL LINERs, in the high state. The upper and lower contours representing  log $L_{\rm{H}\alpha}$  vs. log $L_{\rm{[O~III]}}$ measurements of SDSS DR7 quasars and Sy~1 galaxies from \citet{Shen2011} and \citet{Mullaney2013} show that this sample is up to an order of magnitude underluminous in [O~III], due to  light-travel time delays of an extended narrow line region that has yet to respond to the continuum flux change. The lower limit of \podrick~is due to the [O~III] $\lambda$5007 emission line not being resolved in the low-resolution ($R\sim100$) follow-up spectra. Adapted from Figure 6 in \citet{Gezari2017}.}
\label{fig:haoiii}
\end{figure*}

It is evident from all follow-up spectra of \tyrion~in Figure~\ref{fig:tyrion_spc} that it is also missing the prominent Fe~II pseudo-continuum complex characteristic of NLS1s. Therefore we do not utilize an Fe~II template in  subsequent optical nor UV spectral fitting. The intense ionizing radiation and high temperatures inferred from the presence of the coronal line emission should make visible the multiply ionized Fe~II were it present.  
The fact that Fe~II lags behind H$\beta$ in reverberation mapping studies of AGN \citep{Barth2013} could mean that not enough time has passed for this component to be irradiated, consistent with the weak presence of [O~III] (Figure~\ref{fig:haoiii}). 
\citet{Runnoe2016} also found that, for some CLAGN, the Fe~II complex was only present in the ``on'' state. 
In AGN there is a robust negative correlation between [O~III] and Fe~II (the so-called Eigenvector 1; \citet{Boroson1992}), which manifests typically as weak [O~III] in NLS1s (e.g. \citealt{Rakshit2017}), possibly indicating we should expect Fe~II to become stronger in \tyrion~after the light-travel delay time.

Narrow He~II is frequently observed in AGN, however we measure strong He~II broader than the Balmer emission lines (Figure~\ref{fig:fwhm}), possibly revealing an inner nuclear region not typically probed by the Balmer emission lines alone.  
This has been seen in a number of Seyferts such as the Sy1 Mrk 509, but is far less common.

He~II $\lambda$1640 and [C~III] $\lambda$1909 observed in the UV spectrum are consistent with the presence of higher ionization coronal lines in the optical. All prominent emission features are similar in strength and width to those in the {\it HST} FOS spectrum of NLS1 Mrk 335 and Mrk 478, shown in Figure~\ref{fig:hst} for comparison, however, with a Mg II $\lambda$2798, which is only marginally detected in the first {\it HST}/STIS epoch, and then brightens significantly 4 months later. 
However, like [O~III], the late-time brightening of Mg II is likely a result of light travel time delays if the Mg II and [O~III] line emitting gas resides further out from the central black hole.

\subsubsection{The Accretion Rate of \tyrion}
\label{sec:tyrion_ledd}

The Eddington ratio of \tyrion~ranged between 0.004 and 0.001 from 2018 May to 2019 July, assuming the BH mass of log $M_{\rm{BH}}[M_\odot]$={7.6} (estimate described in Section~\ref{sec:mbh} from stellar velocity dispersion).  Note that we assume a constant for the bolometric correction, but the SED is likely changing throughout the evolution of this source given the dramatic variability in $\alpha_{\rm{OX}}$ described below. This $L/L_{\rm{Edd}}$ is below that of the NLS1 distribution, and on the high end for that of {the low state} CLQs \citep{Xu2012,MacLeod2019}. The range of Eddington ratios for the remainder of the sample is {0.002$-$0.06}. 
\tyrion~is probing a critical {region} in $\alpha_{\rm{OX}}$ and Eddington ratio space related to accretion rate driven state changes analogous to that of X-ray binaries \citep{Ruan2019}.

\subsubsection{X-ray Light Curve and Spectra of \tyrion}
\label{sec:x-ray}

We initially measure a soft X-ray luminosity of a few $\times 10^{41}$ erg s$^{-1}$ from the first {\it Swift} XRT observations of \tyrion~on 2018 July 30. \citet{Wang2011} require at least a few $\times 10^{42}$ erg s$^{-1}$ to power the CLR, a level which \tyrion~did not reach until $\sim$40 days later.  
The XRT light curve in the lower panel of Figure~\ref{fig:xrt} shows that \tyrion~is a variable X-ray source (we note that high-amplitude X-ray variability is characteristic of NLS1s; e.g. \citealt{Nikolajuk2009}). The excess variance (or fractional amplitude of variability) defined by \citet{Nandra1997} as $\sigma_{\rm{rms}}^2=\frac{1}{N\bar{x}}\sum_{i=1}^{N} (x_i-\bar{x})^2 - \delta x^2$ of the 0.3-10 keV 130-day light curve\footnote{The detections used to compute the excess variance were in units of counts.} is 0.41, similar to that of the most variable NLS1s, but high for Sy~1s \citep{Grupe2000}. 
We measure a maximum luminosity of $L_X=(3.7\pm0.4)\times 10^{42} $ erg s$^{-1}$. 
This X-ray luminosity is difficult to obtain with even the brightest supernova explosions, which have been observed up to $\sim10^{41}$ erg s$^{-1}$ \citep{Immler2003}, and it is toward the lower end for both Seyferts and NLS1s \citep{Hasinger2008}. 
The hardness ratio light curve in the lower panel of Figure~\ref{fig:xrt} shows that the X-ray flare is primarily in the soft band i.e. 0.3$-$1.5 keV, while the 1.5$-$10 keV light curve tracks the variability but with a much smaller amplitude. In contrast, the optical and UV photometry displays a plateau during this time, reminiscent of that of \bco~ (Figure~\ref{fig:lc}, \ref{fig:16bco}), before declining over several months in a manner similar to \podrick. 


The simultaneous optical-to-X-ray spectral slope ratio ($\alpha_{\rm{OX}}$) defined as 
\begin{equation*}
\alpha_{\text{OX}}= 0.3838~\text{log}(
L_{\text{2 keV}}/L_{2500\rm{A}})
\end{equation*}
by Eq. 4 of \citet{Tananbaum1979}, and Eq. 11 of \citet{Grupe2010}, over several epochs, measures roughly how an object's SED is changing with time, and is strongly correlated with Eddington ratio \citep{Poole2008}. However, \citet{Grupe2010} argue that 
this correlation is only a reliable estimator for Eddington ratio for sources with $\Gamma \leq 1.6$, above which the relationship saturates.  We derive $\alpha_{\rm OX}$ from the multi-epoch concurrent observations between 2018 July 30 (61 days after discovery) and Dec 08 by {\it Swift} XRT and UVOT (taken with the UVW2 filter, which has a central wavelength of 1928\AA~and FWHM 657\AA; \citet{Poole2008}). 

\tyrion~shows dramatic variability in the X-rays (rising by an order of magnitude in 5 months with $L_X$ that varied between (0.4$-$ 3.1) $\times 10^{42} $ erg s$^{-1}$; see Figure~\ref{fig:xrt}). However, the range of $\alpha_{\rm{OX}}$ for \tyrion~{in its ``on'' state (-1.91$-$-1.40; values listed in Table~\ref{tab:xray}}) is consistent with that of Type 1 Seyferts ($-2.0 < \alpha_{\rm{OX}} < -1.2$;  \citet{Steffen2006,Elvis1994}) and most NLS1s ($-1.8 < \alpha_{\rm{OX}} < -0.9$; \citet{Gallo2006} and systematically steeper than that of typical LINER values ($-1.4 < \alpha_{\rm{OX}} < -0.8$; \citet{Maoz2007}). 




The soft X-ray spectrum and coronal line emission in \tyrion~are shared characteristics with NLS1s.  The soft X-ray component in excess above the extrapolation of hard X-ray power-law continuum is observed in a large fraction of Seyfert AGN \citep{Singh1985}, but is particularly strong in NLS1s. 
The full extent of the soft excess component remains unknown, and its origin is debated. 
It has been ruled out as the tail of the UV thermal emission from the accretion disk \citep{Gierlinski2004,Porquet2004,Piconcelli2005,Miniutti2009} but Comptonization of those seed photons by an optically thick medium is now one of the favored scenarios (e.g. \citealt{Done2012}),
as is blurred ionized disk reflection \citet{Garcia2019}.

Due to their high ionization potentials ($\chi~>$ 100 eV), coronal lines can probe the soft X-ray excess indirectly, as well as the SED in the vicinity of 200 eV, which is difficult to observe otherwise because of both Galactic and intrinsic photoelectric absorption, but important due to their significant contribution to $L_{\rm{bol}}$. 
\citet{Erkens1997} found that coronal lines were more likely to be present in Seyferts with steeper X-ray spectra.
\citet{Gelbord2009} found in a sample of Seyfert galaxies a correlation between soft X-rays and [Fe~VII], [Fe~X], and [Fe~XI] lines, proposed by \citet{Murayama1998a,Murayama1998b} to originate from the innermost wall of the dusty torus (see also \citet{RodriguezArdila2002}). 

NLS1s also display strong coronal line emission (e.g. \citealt{Stephens1989}). 
Optical coronal lines include the forbidden transitions of iron, [Fe~XIV] $\lambda$5304, [Fe~VII] $\lambda$6088, [Fe~X] $\lambda$6376 and [Fe~XI] $\lambda$7894, as well as [Ar~XIV] $\lambda$4414 and [S~XII] $\lambda$7612. 
The coronal lines in NLS1s can be blueshifted with asymmetric velocity profiles and broad wings, consistent with an outflow 
\citep{Erkens1997,Porquet1999,Nagao2000}. 
\citet{Gelbord2009} found [Fe~X]/[O~III] to be the most extreme (by a factor of 2-3) in NLS1s with the narrowest broad lines (FWHM(H$\alpha$)$\sim$800 km s$^{-1}$) during a search for AGN with strong coronal lines in SDSS, and interpreted these sources as having strong soft excesses.

\section{Discussion} \label{sec:disc}
While the number of CLAGN is steadily increasing, there has yet to be a large-scale systematic study of newly-discovered candidates that simultaneously tracks the appearance of continuum variability and the broad-line emission in real-time using high-cadence difference imaging photometry. 

The best-studied target-of-interest in this sample was identified from ZTF based on its TDE-like rise time, and therefore we obtained several epochs of supporting data in real-time throughout its evolution. Its months-long plateau, UV/optical spectra, and high-energy properties were indicative of having changed look to a NLS1. 
Although they are typically highly X-ray variable, such dramatic optical variability of a NLS1 has only been seen in {seven} other sources to-date, including CLAGN NGC 4051 \citep{Guainazzi1998,Uttley1999}, {and SDSS J123359.12+084211.5 \citep{MacLeod2019},} although {they both} changed from an obscured Sy~2 and not a LINER\footnote{{ The remaining objects are CSS100217:102913+404220 \citep{Drake2011}, ULIRG F01004-2237 \citep{Tadhunter2017}, PS16dtm \citep{Blanchard2017}, OGLE17aaj \citep{Gromadzki2019}, , and AT2017bgt \citep{Trakhtenbrot2019}, all of which are discussed further in Section~\ref{sec:t_vs_a}.}}. \tyrion~is therefore unique not only among this sample, but among CLAGN overall.

\subsection{A New Class of changing-look LINERs}

We establish this particular class of CLAGN associated with extreme order-of-magnitude changes in continuum and emission line flux compared to less dramatic changing looks occurring in Seyferts (shown in Figure~\ref{fig:macleod}).

Although most CLAGN reported to-date are Seyferts, this may be due to sample selection bias, as the high numbers of LINERs may cause them to be seen as galaxy contaminants in such searches. Difference imaging offers a unique mechanism to discover variability in known LINERs.

\subsection{Is \tyrion~a TDE or AGN activity?} \label{sec:t_vs_a}

We focus specifically on \tyrion, which shows the appearance of broad Balmer and coronal lines within 16 years of being spectroscopically confirmed as a LINER, accompanied by an order-of-magnitude soft X-ray flare. Given a ROSAT All-Sky Survey flux upper limit of $F_{0.1–2.4~\rm{keV}} < 5\times10^{-13}$ ergs s$^{-1}$ cm$^{-2}$ at the location of the host from 1990 to 1991 \citep{Voges1999}, \tyrion~has therefore displayed a changing look in both the optical and X-ray usages of this term. 
The lower limit for this change in soft X-ray flux (0.1-2.4 keV) was by a factor of 7 at the time of the most recent observation. 

Although highly photometrically variable on their own, flares due to non-AGN mechanisms are not unheard of in NLS1s. 
For example, CSS100217:102913+404220 displayed a high state ($M_V = -22.7$ at 45 days post-peak) accompanied by broad H$\alpha$ and was interpreted either as a Type IIn SN \citep{Drake2011} or TDE \citep{Saxton2018} near the nucleus ($\sim$150 pc) of a NLS1. It eventually faded back to its original level after one year. 
PS16dtm (or iPTF16ezh/SN 2016ezh) was a $\sim1.7\times10^{4}$ K, and near-Eddington but X-ray-quiet nuclear transient with strong Fe~II emission which plateaued over $\sim$100 days while maintaining a constant blackbody temperature. The event was interpreted as a TDE exciting the BLR in a NLS1 \citep{Blanchard2017}, although \citet{Oknyansky2018a} claimed it may instead be a CLAGN transitioning into a Sy~1. No X-rays were observed during follow-up, dimming at least by an order of magnitude compared to archival observations, but were predicted to reappear after the obscuring debris had dissipated. 
SDSS J1233+0842 was discovered as a CLQ when it changed into a composite type galaxy or transition object (with [O~III]/H$\beta=-0.10$ and [N~II]/H$\alpha=-0.17$ from Figure 2.a. in \citet{MacLeod2019}). It shows variable Fe~II emission (similar to PS16dtm), with the broad line emission disappearing between 2005 and 2016. 

A nuclear transient in the nearby ULIRG F01004-2237 was classified as a TDE\textemdash despite an unusually long peak time of 1 year\textemdash partially  based on the strength of its He~II compared to H$\beta$ \citep{Tadhunter2017}. This ratio was unprecedented for AGN activity, even for AGN in the high state of a changing look. We note that although it is broad, He~II/H$\beta \sim$ 0.4 for \tyrion~is far below that measured for F01004-2237. It was later argued that the nature of this transient may instead be due to changes in the accretion flow, similar to that of OGLE17aaj, which also showed a slow optical rise and long plateau and slow decline and UV and X-ray properties similar to that of \tyrion, although it lacked spectral classification prior to discovery of the transient \citep{Gromadzki2019}. 
The transient AT2017bgt was classified as a dramatic SMBH UV/optical flare which irradiated the BLR and was interpreted as the result of increased accretion onto the SMBH \citep{Trakhtenbrot2019}. Unlike \tyrion, it showed no decrease in flux over several months. The persistence of the UV emission distinguished it from SNe and TDEs, and the extremely intense nature of the UV continuum as well as presence of Bowen fluorescence He~II, [N~III] $\lambda$4640, and [O~III] double-peaked features in the unobscured optical spectrum distinguished it from CLAGN. As in the ``on'' state of \tyrion, the Balmer FWHM in all 3 sources are consistent with that of NLS1 galaxies.

ECLEs are most typically thought to be the echoes of TDEs via the accretion of tidal disruption streams by previously non-active SMBHs \citep{Wang2012}. 
However, less than 10 ECLEs have been reported in the literature, most notably SDSS J0952+2143  (\citet{Komossa2008,Komossa2009,Palaversa2016};  also technically a NLS1 using the unconventional cutoff in \citet{Rakshit2017}, see Section~\ref{sec:nls1} for details), and SDSS J0748+4712 \citep{Wang2011}. 
 We confirm that \tyrion~is technically an ``extreme'' CLE by the definition put forth by \citet{Wang2012}, because the strength of [Fe~X] $\lambda$6376 is comparable to that of [O~III] $\lambda$5007, as well as by the presence of [Fe~XIV] in the optical spectrum (seen in Figures~\ref{fig:tyrion_spc} and~21) following the independent definition of \citet{Palaversa2016}. We note, however, that it is the present weakness of [O~III] that is driving this diagnostic, and the coronal lines overall do not appear nearly as strong when compared to the prototypical ECLEs, SDSS J0952+2143 and J0748+4712, in Figure~\ref{fig:tyrion_spc}. 
 This strong, slowly variable transient nuclear coronal line emission necessitates soft X-ray flaring outbursts from an accretion disk, which may be formed as tidal debris settles, illuminating the outermost debris as well as intervening ISM \citep{Komossa1999}.  The coronal lines in these sources, some blueshifted, faded on timescales of 1-5 years, with strong [O~III] appearing even later. Because strong coronal line emission is not a TDE diagnostic in isolation, some ECLE galaxies with persistent coronal lines may instead be Seyferts. 

IC 3599 is an optical changing-look (displaying dramatic variability in not only Balmer lines but also [Fe~VII] and [Fe~XIV]) Sy~1.9  galaxy with strong soft X-ray repeating outbursts from its galactic nucleus which can be modeled by a disk instability with a rise time of $\sim$1 year whereby the inner disk is vacated and subsequently refills \citep{Brandt1995,Grupe1995,Komossa1999,Campana2015,Grupe2015}. It is the only AGN which has shown fading of its coronal lines (though this variability is common among non-active ECLEs). 

The {\it Swift}/XRT and {\it XMM} spectra of \tyrion~fit well to a steep power law ($\Gamma\sim$3$\pm$0.2) below 2 keV, not a disk blackbody as would be expected in the TDE scenario (see Figures~\ref{fig:xrt} and \ref{fig:xmm}). 
{Fitting the higher signal-to-noise XRT data to a blackbody+power law with Galactic absorption worsened the fit significantly ($\chi^2=227.5/247$ compared to $\chi^2=171.00/245$ for a simple absorbed power law).}
The large covering factor measured for \tyrion~by {\it Spitzer} is also more consistent with mid-infrared studies of CLAGN \citep{Sheng2017}, than the covering factor derived for TDEs with dust echoes (with $f_{\rm{dust}}=E_{\rm{dust}}/E_{\rm{absorb}}$ at the $\sim$1\% level; \citealt{vanVelzen2016}).  This could imply appreciable accretion happening recently, because that is very likely required for a dusty torus with a large covering factor.  In an accretion event unrelated to disk physics, a self-gravitating molecular cloud with low enough angular momentum could also be efficiently accreted on the correct timescales, activating radiation which subsequently illuminates the BLR (e.g. \citealt{Hopkins2006}).  One way to obtain a larger covering factor would also be via chaotic cold accretion, by which interaction via inelastic collisions is made easier, boosting the funneling of molecular clumpy clouds toward the SMBH, and therefore enhancing the accretion rate \citep{Gaspari2017}. 
The high blackbody temperature measured from UV spectroscopy implies the line of sight to the transient is not significantly dust obscured. 
\citet{Sheng2017} argue that mid IR light echoes of CLAGN (with $\Delta W1 | W2 \gtrsim 0.4$ mag)  was additional evidence to support the reprocessing scenario driven by changing accretion rate instead of variable obscuration. $W1 - W2$ for that sample varied between 0.1 and 1.2 mag, so $[3.6]-[4.5]~\mu m=1.4$ mag for \tyrion~was consistent with the lowest end of that sample for mid IR color (it would not have been selected based on its variability amplitude for the short duration of the {\it Spitzer} observations reported here).


LINERs may have inefficient accretion disks surrounding a low-luminosity AGN, occupying a unique physical parameter space compared to other CLAGN. 
Similar to the unification scheme derived for AGN \citep{Antonucci1993,Urry1995}, broad- and narrow-line LINERs can be categorized into LINER1s and LINER2s (e.g. \citealt{Ho1997a,Ho1997e,GonzalezMartin2015}). 
\citet{Yan2019} reported the discovery of the ``turning on'' of a type 1 Seyfert occurring in LINER SDSS1115+0544~which flared for $\sim$1 year and subsequently plateaued, followed by a mid-IR dust echo delayed with respect to the optical by 180 days and a late-time UV flare, although no soft X-rays were detected then. Narrow coronal lines appeared in the spectrum along with H$\alpha$ and H$\beta$ consistent with broad line emission. As was done in \citet{Yan2019}, we measured the soft X-ray-[Fe~VII] ratio for \tyrion~to be  log $L_{2~\rm{keV}}/L_{\rm{[Fe~VII]\lambda6088}}$ =1.25 at maximum, still significantly below the average of 3.37 and pointing to an X-ray deficit compared to normal AGN \citep{Gelbord2009}, although we note that the soft X-rays changed by a factor of 10 and likely continued to rise beyond our last {\it Swift} observation.  
We also measure a minimum $L/L_{\rm{Edd}}$ equivalent to that of SDSS1115+0544. 
\citet{Yan2019} concluded an instability was required to ``turn on'' an AGN from a quiescent galaxy within hundreds of days. They argued that (despite a rate in tension with the AGN duty cycle) given the discovery of \bco~ and SDSS1115+0544 one year apart, such events should not be uncommon, a prediction this sample supports. There must be a connection between the LINER hosts and the state that is enabling these rapid transitions. 


\subsection{The nature of the high-ionization forbidden ``coronal'' lines in \tyrion} 
\citet{Noda2018} posited that in the well-studied changing-look AGN Mrk 1018, the coming and going of the soft X-ray excess (the main ionization source) drives the appearance and disappearance of the BLR and therefore the changing-look phenomenon. 
We observe strong soft X-rays increasing in luminosity over time, which are required to form the coronal lines, although we note that the peak of the X-ray flaring appears to lag behind the UV/optical flaring. 

The nuclear outburst in UV and X-ray required is similar to cataclysmic variable or black hole binary thermal-viscous disk instability flares, which have been discussed as a possible mechanism for powering optical changing-looks, although the observed time scales are much faster than predicted (e.g. \citealt{Siemiginowska1996,Lawrence2018,Stern2018,Ross2018}). 

\citet{Ross2018} attribute changing looks to a thermal (cooling) front propagating inward through the accretion disk or disk surface opacity changes, which have the correct timescales for observed transitions, unlike other proposed CLAGN mechanisms. 

We posit that this quiet LINER suddenly goes into an active outbursting state, the rise in ionizing radiation at first confined to the innermost BLR, turning on into a NLS1, then flash ionizing the ambient gas in the CLR, whereas the NLR (where [O~III] and Fe~II lines are formed) is at larger distances, and thus light-travel time effects delay their response.  Mg II, though still broad, is formed further out on average \citep{Goad1993,OBrien1995,Cackett2015}.

\subsection{The nature of the soft X-ray excess during the NLS1 state of \tyrion}

The preceding interpretation does not explain the soft X-ray rise, which is clearly delayed at least $\sim$60 days with respect to the end of the UV/optical rise (shown in Figure~\ref{fig:spitz}), and may speak instead to a lag in an ``outside-in'' sense following the direction of an accretion flow, rather than photon propagation from a central ``lamp post''. 
This is in contrast to the clear inter-band time lags on the order of days in support of the reprocessing scenario measured by \citet{Shappee2014} in high cadence multiwavelength observations of CLAGN NGC 2617, which transitioned from a Sy~1.8 to a Sy~1 in 10 years. The $\sim$2 month lag observed in \tyrion~also suggests that this delay is not simply from light-travel time. 
X-ray inter-band time delays in NLS1s measured via Fourier based spectral timing, due to either X-ray reverberation or propagating fluctuations, are typically on the order of tens to hundreds of seconds (e.g. \citealt{Uttley2014,Kara2016}).

This delayed X-ray response may tell us something fundamental about the origin of the soft X-ray excess in AGN in general. The long delay of the soft X-ray flare relative to the expected light-travel time delays between the UV/optical emitting accretion disk and the compact, hot corona suggests 
that we are witnessing the real-time assembly of the corona plasma itself, possibly due to structural changes due to the dramatic change of state in the inner accretion disk \citep{Garcia2019}. 

If the Balmer emission is indeed from a BLR, we predict the H$\alpha$ and H$\beta$ lines should get broader as the UV luminosity decreases.
Continued spectroscopic monitoring to look for evolution in line widths and strengths, particularly the narrow [O~III] emission line and Mg II, and monitoring of the soft X-rays  will be critical to map out the structure of this system and distinguish between the scenarios presented here.

\section{Conclusions} \label{sec:concl}
We present the changing looks of \num~known LINERs caught turning on into type-1-like AGN found in Year 1 of the ZTF survey. It is the first systematic study of its kind performed in real time using difference imaging variability as the discovery mechanism for selecting nuclear transients in these previously quiescent galaxies.
\begin{easylist}[enumerate]
	& We establish a class of changing-look LINERs, distinct from Seyfert \clagn, with unique spectroscopic and photometric variability properties intrinsically due to the LINER accretion state.
	& In their "on state" the changing-look LINERs have suppressed narrow [O~III] line emission compared to normal AGN of the same broad H$\alpha$ luminosity, and inferred Eddington ratios 1$-${3} orders of magnitude above their LINER state.
	& This sample includes a multiwavelength study between 2018 June to 2019 March of the first case of a LINER changing look to a NLS1 --- \tyrion~--- which transitioned within 3 months based on its archival light curve.
	& We observed the delayed response of the NLR and broad Mg II with respect to the appearance of broad (yet $< 2000$ km s$^{-1}$) Balmer lines, and X-ray flaring delayed $\sim$60 days with respect to the optical/UV rise of this nuclear transient, indicative of an ``outside-in'' transition. 
	& We interpret this particular object to be a dramatic change of state in a pre-existing LINER accretion disk, which eventually forms an optically thick inner structure that up-scatters the UV/optical seed photons to produce a delayed soft X-ray excess. \newline 
\end{easylist}
This class of previously-weak AGN has the  potential to be a laboratory with which to map out the structure of the accretion flow and surrounding environment. We plan to continue to monitor the behavior of these transients, and expect to build upon the sample at a rate of $\sim$4 year$^{-1}$ for the next two years of the ZTF survey.

\acknowledgments
{\bf Acknowledgments}  ---
{We thank the referee for insightful and constructive comments and suggestions.} S.G. acknowledges E. Quataert, and S.F. acknowledges R. Mushotzky for useful discussions {and J. Ruan for helpful correspondence}. We would like to thank S.M. Adams, M. Kuhn, and Y. Sharma for obtaining the Keck and Palomar 200-inch spectral observations.  S.G. is supported in part by NSF CAREER grant 1454816, and the XMM-Newton grant for AO-17 Proposal 82204.

Based on observations obtained with the Samuel Oschin Telescope 48-inch and the 60-inch Telescope at the Palomar Observatory as part of the Zwicky Transient Facility project. ZTF is supported by the National Science Foundation under Grant No. AST-1440341 and a collaboration including Caltech, IPAC, the Weizmann Institute for Science, the Oskar Klein Center at Stockholm University, the University of Maryland, the University of Washington, Deutsches Elektronen-Synchrotron and Humboldt University, Los Alamos National Laboratories, the TANGO Consortium of Taiwan, the University of Wisconsin at Milwaukee, and Lawrence Berkeley National Laboratories. Operations are conducted by COO, IPAC, and UW.

SED Machine is based upon work supported by the National Science Foundation under Grant No. 1106171.

This work was supported by the GROWTH project funded by the National Science Foundation under Grant No 1545949.

These results made use of the Discovery Channel Telescope at Lowell Observatory.
Lowell is a private, non-profit institution dedicated to astrophysical research and public appreciation of
astronomy and operates the DCT in partnership with Boston University, the University of Maryland, the
University of Toledo, Northern Arizona University and Yale University. The upgrade of the DeVeny optical spectrograph has been funded by a generous grant from John and Ginger Giovale.

The W. M. Keck Observatory is operated as a scientific partnership among the California Institute of Technology, the University of California, and NASA; the Observatory was made possible by the generous financial support of the W. M. Keck Foundation.

Based on observations obtained at the Gemini Observatory, which is operated by the Association of Universities for Research in Astronomy (AURA) under a cooperative agreement with the NSF on behalf of the Gemini partnership: the National Science Foundation (United States), the National Research Council (Canada), CONICYT (Chile), the Australian Research Council (Australia), Ministerio da Ciencia e Tecnologia (Brazil) and Ministerio de Ciencia, Tecnologia e Innovacion Productiva (Argentina).

Based on observations made with the NASA/ESA Hubble Space Telescope, obtained from the data archive at the Space Telescope Science Institute. STScI is operated by the Association of Universities for Research in Astronomy, Inc. under NASA contract NAS 5-26555.  These observations are associated with program 15331.

This work made use of data supplied by the UK Swift Science Data Centre at the University of Leicester.

This research has made use of the NASA/IPAC Extragalactic Database (NED) which is operated by the Jet Propulsion Laboratory, California Institute of Technology, under contract with the National Aeronautics and Space Administration.

\facilities{PO:1.2m, PO:1.5m, Hale, DCT, Keck:I (LRIS), Gemini:Gillett, HST (STIS), Swift, XMM}

\software{
HEAsoft \citep{Arnaud1996},
SAS \citep{Gabriel2004}
}

\bibliographystyle{aasjournal}

\bibliography{CLAGN}

\end{document}